\documentclass[11pt]{article}

\usepackage[T1]{fontenc}
\usepackage[utf8]{inputenc}
\usepackage{lmodern}
\usepackage[a4paper,margin=1in]{geometry}

\usepackage{amsmath,amssymb}
\usepackage{mathtools}
\usepackage{tabularx}
\usepackage{graphicx}
\usepackage{float}          
\usepackage{subcaption}     
\usepackage{booktabs}       
\usepackage{xcolor}
\usepackage{authblk}        

\usepackage[hidelinks]{hyperref}
\usepackage[capitalise]{cleveref}

\usepackage[backend=biber,style=numeric-comp,sorting=none,url=false,eprint=false,maxbibnames=10]{biblatex}
\newcommand{\ethics}[1]{\paragraph{Ethics.}~#1}
\newcommand{\dataccess}[1]{\paragraph{Data accessibility.}~#1}
\newcommand{\aucontribute}[1]{\paragraph{Authors' contributions.}~#1}
\newcommand{\competing}[1]{\paragraph{Competing interests.}~#1}
\newcommand{\funding}[1]{\paragraph{Funding.}~#1}

\title{\bfseries Mathematical modelling of immune persistence and relapse pathways in CAR T-cell therapy for B-ALL}

\author[1,2]{Alexis Farman\thanks{Corresponding author: \texttt{alexis.farman.23@ucl.ac.uk}}}
\author[1,2]{Benjamin J. Walker}
\author[3]{Martin A. Pule}
\author[1,2]{Karen M. Page}
\affil[1]{Department of Mathematics, University College London, United Kingdom}
\affil[2]{Institute for the Physics of Living Systems, University College London, United Kingdom}
\affil[3]{Cancer Institute, University College London, United Kingdom}
\date{}

\begin{document}

\maketitle

\begin{abstract}
Chimeric antigen receptor (CAR) T-cell therapy has transformed the treatment of B-cell acute lymphoblastic leukaemia (B-ALL). Despite high initial response rates, a substantial fraction of patients relapse, often due to loss of CAR T-cell persistence, antigen escape, or immune-privileged sites that shield tumour cells. Prolonged CAR T-cell persistence is clinically associated with durable remission, but why it is required remains poorly understood. To address this, we develop and analyse the BEAM (Blast, Effector, Activated, Memory) model of CAR T-cell dynamics in B-ALL. BEAM extends predator--prey models with three CAR T-cell states (memory, activated, effector) coupled to a logistic growth equation for the blasts, calibrated against the FELIX trial of obecabtagene autoleucel in adult B-ALL. We find that both memory and effector persistence prevent relapse, but for distinct reasons: memory persistence sustains surveillance against low-burden or slowly proliferating residual disease, while effector persistence clears isolated blasts emerging from immune-privileged sites. The model further predicts a trade-off between immediate cytotoxicity and durable surveillance, and identifies initial tumour burden as a key modifiable factor for reducing antigen-negative relapse. Together, these results offer a framework for designing more durable, individually tailored CAR T-cell therapies.
\end{abstract}


\bigskip
\section{Introduction}

CAR T-cell therapy genetically engineers a patient's own T-cells to recognise and kill cancer cells, and can induce deep, durable remissions in otherwise refractory haematologic cancers \cite{neelapu2017axicabtagene, schuster2019tisagenlecleucel}. In relapsed or refractory B-cell acute lymphoblastic leukaemia (B-ALL), clinical trials of anti-CD19 CAR T-cells have reported remission rates exceeding 80\% \cite{maude2018tisagenlecleucel, roddie2024obecabtagene}. However, 10–20\% of patients fail to respond to therapy and, among those achieving remission, up to half relapse within the first year \cite{maude2018tisagenlecleucel, park2018long, roddie2024obecabtagene}. Understanding the mechanisms underlying tumour relapse is critical to improving CAR T therapy.

Several biological processes have been implicated in CAR T-cell treatment failure. Antigen escape accounts for 10–30\% of relapses \cite{sotillo2015convergence, majzner2018tumor, ruella2016dual}. Leukaemic cells may also evade detection by occupying immunoprivileged bone marrow niches \cite{witkowski2020extensive, tettamanti2022catch}. Perhaps most critically, the persistence of CAR T-cells themselves appears to be a determinant of durable remission, whereas relapse can coincide with loss or contraction of the CAR T-cell population \cite{park2018long, stein2019tisagenlecleucel, melenhorst2022decade, roddie2024obecabtagene}.Yet why long-term persistence is required, and which CAR T-cell characteristics promote this persistence remains incompletely understood~\cite{shah2019mechanisms, sterner2021car}.


Two mechanistic explanations are plausible: residual disease may rebound from very low cell numbers and require sustained effector surveillance to be cleared, or relapse may be due to the eventual depletion of the memory pool through repeated activation. The two pathways have distinct therapeutic implications: one points to engineering longer-lived effectors, the other to memory-biased products or repeated dosing.

Mathematical modelling can relate treatment outcome to quantities that are hard to disentangle clinically, such as cell dose, tumour growth and antigen expression. Classical models of tumour–immune dynamics have used predator–prey ordinary differential equation (ODE) formulations to capture essential features of tumour growth and immune-mediated killing \cite{eftimie2011interactions, de2005validated, frascoli2014dynamical}. More recently, several authors have adapted such frameworks specifically to CAR T-cell therapy. \citeauthor{sahoo2020mathematical} developed the CARRGO model to study CAR T-cell proliferation and exhaustion in glioblastoma. More complicated models have incorporated cytokine dynamics to model the onset and progression of cytokine release syndrome (CRS) \cite{hardiansyah2019quantitative}. Other studies have included additional immune cell populations \cite{owens2021modeling, kimmel2021roles, perez2021car} or stratified tumour compartments by antigen expression to investigate antigen-negative escape \cite{liu2022computational, santurio2024mechanisms}.

Others have modelled CAR T-cell kinetics and phenotype directly, using cellular kinetic models of expansion, contraction and persistence \cite{stein2019tisagenlecleucel} and quantitative systems pharmacology models that resolve effector and memory populations \cite{hardiansyah2019quantitative, mueller2021early}. Further models represent memory-like compartments explicitly and link them to response duration \cite{martinez2021mathematical, barros2021cart, mueller2021early}, or describe antigen-driven transitions among memory, effector and exhausted states \cite{kirouac2023deconvolution}. 

Despite this progress, several important gaps remain. First, few existing models have quantitatively separated the contributions of memory and effector CAR T-cell persistence to long-term tumour control. Second, the interaction between immune persistence and the multiple relapse routes observed clinically (loss of CAR T-cell persistence, antigen-negative escape, and temporary protection of blasts in immune-privileged niches) has not been systematically compared within a single framework. Third, most existing models focus on cellular kinetic fitting, early response, toxicity, or deterministic tumour-control dynamics, and do not explicitly distinguish deterministic dormancy from stochastic elimination when blast numbers become very small.

In this paper, we develop the BEAM (Blast, Effector, Activated, Memory) model to address these questions. Building on classical predator--prey systems, BEAM distinguishes three CAR T-cell functional states (memory, activated, and effector) in their interactions with proliferating leukaemic blasts, allowing the contributions of memory and effector persistence to long-term tumour control to be examined separately. Model parameters are informed by clinical trial data of obecabtagene autoleucel (obe-cel; a 4-1BB$\zeta$ anti-CD19 CAR T-cell product) for adult B-ALL from the FELIX study~\cite{roddie2024obecabtagene}, supplemented by experimental measurements from the literature. The main contribution is the combination of a memory--activation--effector structure with deterministic stability analysis and a hybrid deterministic--stochastic simulation framework, which together allow us to analyse deterministic dormancy states, resolve stochastic extinction of very small residual blast populations, and model loss of target antigen events within a single framework.

\section{Methods} \label{methods}
\subsection{Model equations} \label{modeldev}

\begin{figure}[h!]
    \centering
    \includegraphics[trim={4.8cm 4.2cm 8cm 2cm},clip, width=.8\linewidth]{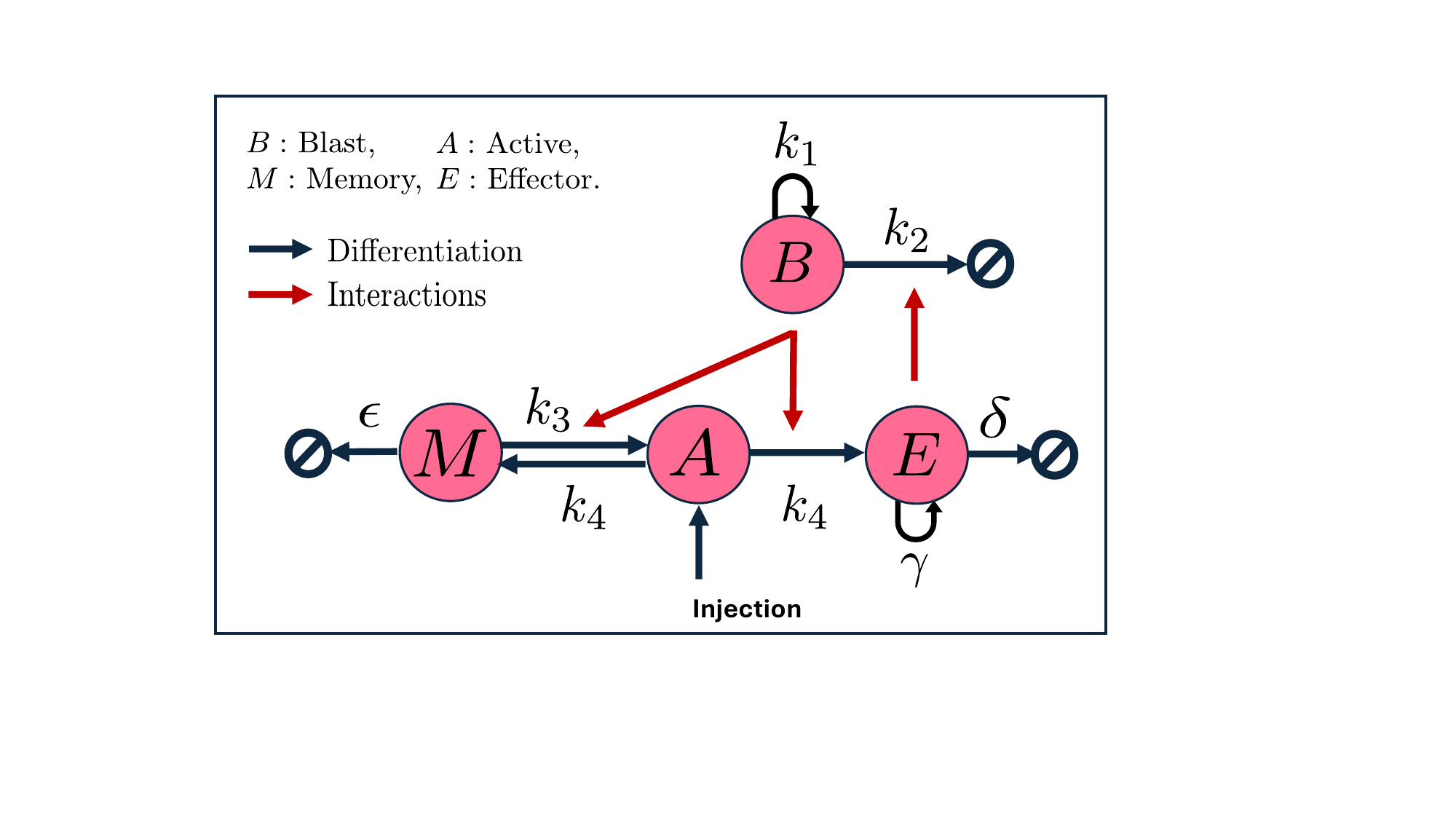}
    \caption{Schematic of the BEAM model (Equations (\ref{eq_blast_1})-(\ref{memory1})). The circles represent cell populations (memory $M(t)$, activated $A(t)$, effector CAR T-cells $E(t)=\sum_{i=1}^{N}E_i(t)$ and leukaemic blasts $B(t)$). Differentiations, births and deaths are indicated by black lines while interactions are illustrated by red lines. The injection arrow denotes CAR T-cell infusion into the phenotypic compartments specified by the initial conditions, rather than assuming that all infused cells are activated at $t=0$.}
    \label{fig:c2_model_schematic}
\end{figure}
The BEAM (Blast, Effector, Activated, Memory) mathematical model builds upon classical predator--prey type ODE models of tumour--immune interactions. Our model tracks leukaemic blast cells $B$ and three CAR T-cell functional states: resting memory cells ($M$), activated cells ($A$), and effector cells, which are subdivided into $N$ division compartments $E_i$, $i=1,\ldots,N$. We write the total effector population as $E(t)=\sum_{i=1}^{N}E_i(t)$. The number of leukaemic blast cells $B$ is governed by
\begin{equation} \label{eq_blast_1}
    \underbrace{\frac{\mathrm{d}B}{\mathrm{d}t}}_{{\substack{\text{cancer cells} \\ \text{rate of change}}}}
    = \underbrace{k_1 B \left(1-\frac{B}{K} \right)}_{{\substack{\text{logistic growth} \\ \text{of cancer cells}}}}  -  \underbrace{k_2 B \sum_{i=1}^{N} E_i}_{{\substack{\text{CAR T-cell
induced} \\ \text{cancer cell death}}}}. 
\end{equation}
Blast cells proliferate at rate $k_1$ up to a carrying capacity $K$. $E_i$ is the number of effector CAR T-cells in division compartment $i$, and $k_2$ is a lumped tumour-killing parameter capturing CAR T-cell encounter, recognition and cytotoxic activity. The effector CAR T-cells are assumed to go through $N=6$ division stages (see \Cref{tab:ndparamc2}; calibrated against FELIX kinetics), with their numbers governed by
\begin{subequations} \label{effector1}
  \begin{align}
    \overbrace{\frac{\mathrm{d}E_1}{\mathrm{d}t}}^{{\substack{\text{rate} \\ \text{of change}}}} \;\;\;\;  &= \;\;\;\; \overbrace{2k_4A\frac{B}{B_{1/2} + B }}^{{\substack{\text{effector CAR T-cells} \\ \text{sourced from activated cells}}}} \;\;\;\; - \;\;\;\;\overbrace{\gamma E_1}^{{\text{division}}} ,  \\
    \frac{\mathrm{d}E_i}{\mathrm{d}t}\;\;\;\;  &=\;\;\;\; \underbrace{\gamma (2 E_{i-1} - E_{i})}_{{\text{division}}}, \qquad i=2 \hdots N-1, \\
    \frac{\mathrm{d}E_N}{\mathrm{d}t}\;\;\;\; &= \;\;\;\;\underbrace{2\gamma E_{N-1}}_{{\text{division}}} \;\;\;\; - \;\;\;\;\underbrace{ \delta E_N}_{{\text{death}}},
  \end{align}
\end{subequations}
where $\gamma$ is the division rate of the effector CAR T-cells and $\delta$ is the death rate of cells in the final effector compartment. When an effector cell in compartment $E_i$ divides, it produces two daughter cells in compartment $E_{i+1}$. Effector cells are produced by activated CAR T-cells. The activated and memory compartments are governed by
\begin{subequations} \label{memory1}
\begin{align}
    \overbrace{\frac{\mathrm{d}A}{\mathrm{d}t}}^{{\substack{\text{rate} \\ \text{of change}}}} = &\underbrace{k_3 MB}_{{\substack{\text{activation}}}} - \underbrace{k_4 A \left( 1 - \frac{B}{B_{1/2} + B }\right)}_{{\substack{\text{division to memory}}}} - 
    \overbrace{k_4 A \frac{B}{B_{1/2} + B }}^{{\substack{\text{division to effector}}}} , \\
    \frac{\mathrm{d}M}{\mathrm{d}t} = - &\overbrace{k_3 MB}^{{\substack{\text{activation}}}} + \overbrace{2k_4 A \left( 1 - \frac{B}{B_{1/2} + B }\right)}^{{\substack{\text{division to memory}}}} - \underbrace{\epsilon M}_{{\substack{\text{death}}}},
\end{align}
\end{subequations}
where $k_3$ is the rate at which blast cells activate memory cells, $k_4$ is the rate at which activated cells divide, either to effector or memory phenotypes, according to the value of $B_{1/2}$. Equivalently, an activated cell divides into the effector lineage with tumour-burden-dependent probability $B/(B_{1/2}+B)$ and into the memory lineage with probability $B_{1/2}/(B_{1/2}+B)$. The cellular division of an activated cell results in the production of two memory cells $M$ or two effector cells $E_1$. Finally, $\epsilon$ is the death rate of memory cells. Differentiation and interaction pathways between the different cell compartments are summarised with a model schematic in \Cref{fig:c2_model_schematic}. This burden-dependent partitioning is a phenomenological idealisation. It is motivated by experimental evidence that the choice between effector and memory T-cell fate is shaped by cumulative antigenic signalling~\cite{kaech2002effector} and has been used in the modelling literature~\cite{kirouac2023deconvolution}.

The BEAM model treats blast--CAR T-cell dynamics as a well-mixed predator--prey system. Effector CAR T-cells kill blasts following the law of mass-action and progress through $N=6$ divisions at constant rate $\gamma$, producing an expansion peak followed by a contraction consistent with the CAR T-cell kinetics observed clinically~\cite{roddie2024obecabtagene}, before terminal loss at rate $\delta$. Memory cells are activated by mass-action contact with blasts and decay at a small rate $\epsilon \ll \delta$. 

All model variables and parameters are taken to be non-negative. Baseline parameter values are summarised in Table \ref{tab:ndparamc2}, while initial conditions are specified in the parameter-estimation section below. Wherever possible, parameter values were informed by clinical data from the FELIX study of obecabtagene autoleucel (obe-cel; a 4-1BB$\zeta$ anti-CD19 CAR T-cell product) in adult B-ALL \cite{roddie2024obecabtagene}. Because the available clinical data report only total CAR T-cell abundance and aggregate disease burden, we restrict the core model to total blast abundance and three CAR T-cell functional states, with the infused product partitioned into memory-like ($M$, $A$) and effector-like ($E_i$) compartments following the phenotypic breakdown reported in \cite{roddie2024obecabtagene}.

\subsection{Parameter estimation}

Where possible, model parameters are calibrated against FELIX data~\cite{roddie2024obecabtagene}, supplemented by supporting literature. Parameters fall into three groups: those informed directly by FELIX kinetics (effector and memory division and death rates, initial product composition); those estimated from broader T-cell biology with wide plausible ranges (activation rate, activated-cell division rate, phenotypic switching threshold); and the lumped tumour-killing rate $k_2$, which is tuned to place the baseline simulation in the treatment-success regime. Baseline values and ranges are summarised in \Cref{tab:ndparamc2}, with initial conditions in \Cref{tab:initcond}.

\textbf{Tumour dynamics.} The blast proliferation rate $k_1$ is estimated assuming exponential growth of the tumour. The doubling time for B-ALL blast cells is reported to be approximately $T_d=2.5$ days, with a possible range from $1$ to $14$ days \cite{pal2016long,beesley2006authenticity}. This corresponds to a plausible range of approximately $k_1=\ln(2)/T_d \in (0.05,0.7)\;\mathrm{day}^{-1}$. For the baseline simulations, we use $k_1=0.2\;\mathrm{day}^{-1}$, corresponding to a doubling time of approximately $3.5$ days, and explore the wider range in sensitivity analyses. Based on clinical estimates of disease burden, we set a carrying capacity of $K=10^{12}$ cells \cite{saygin2022measurable}.

\textbf{CAR T-cell expansion and persistence.} CAR T-cell kinetic parameters are calibrated against Figure S13 of \cite{roddie2024obecabtagene}, which reports CAR T-cell counts in patients over time. After infusion, CAR T-cells expand roughly 60-fold and peak around day~25. With $N=6$ effector divisions, the choices $\gamma=0.3\;\text{day}^{-1}$ and $\delta=0.2\;\text{day}^{-1}$ reproduce the timing of expansion, peak and contraction over the following three months and are consistent with broader T-cell biology~\cite{kaech2002effector}. FELIX data also suggest memory CAR T-cell numbers halve roughly every two months, yielding $\epsilon=0.01\;\text{day}^{-1}$, within the lifespan range of 10--160 days reported elsewhere~\cite{kaech2002effector}. The lumped tumour-killing rate $k_2$ cannot be identified from FELIX directly. Since the trial dose led to successful treatment in a majority of patients, we set $k_2=4 \times 10^{-10}\;\text{day}^{-1}\,\text{cell}^{-1}$ to place the baseline simulation in the success regime.

\begin{table}[H]
\centering
\caption{Values and ranges for dimensional model parameters with their units and description. Unless stated otherwise, the baseline values are those used in the numerical simulations.}
\label{tab:ndparamc2}
\begin{tabularx}{\textwidth}{@{}cccc>{\raggedright\arraybackslash}Xc@{}}
\toprule
\textbf{parameter} & \textbf{value} & \textbf{range} & \textbf{unit} & \textbf{description} & \textbf{source}\\
\midrule
$k_1$ & $0.2$ & $\left(0.05-0.7\right)$ & day$^{-1}$ & cancer cell growth rate & \cite{pal2016long,beesley2006authenticity}\\
$k_2$ & $4\times10^{-10}$ & $\left(10^{-11}-10^{-9}\right)$ & day$^{-1}$ cell$^{-1}$ & CAR T-cell killing rate & Estimated\\
$k_3$ & $10^{-9}$ & $\left(10^{-10}-10^{-8}\right)$ & day$^{-1}$ cell$^{-1}$ & CAR T-cell activation rate & Estimated, \cite{kaech2002effector}\\
$k_4$ & $0.1$ & $\left(0.01-1\right)$ & day$^{-1}$ & activated CAR T division rate & Estimated, \cite{kaech2002effector}\\
$\gamma$ & $0.3$ & $\left(0.1-0.5\right)$ & day$^{-1}$ & effector CAR T division rate & FELIX, \cite{kaech2002effector,roddie2024obecabtagene}\\
$\delta$ & $0.2$ & $\left(0.1-0.4\right)$ & day$^{-1}$ & effector CAR T death rate & FELIX, \cite{kaech2002effector,roddie2024obecabtagene}\\
$\epsilon$ & $0.01$ & $\left(0.005-0.1\right)$ & day$^{-1}$ & memory CAR T death rate & FELIX, \cite{kaech2002effector,roddie2024obecabtagene}\\
$B_{1/2}$ & $10^{9}$ & $\left(10^{8}-10^{10}\right)$ & cell & threshold blast number for activated CAR T-cell division into effector or memory & Estimated\\
$K$ & $10^{12}$ & & cell & blast carrying capacity & \cite{saygin2022measurable}\\
$N$ & $6$ & & & number of effector divisions & \cite{kaech2002effector,roddie2024obecabtagene}\\
\bottomrule
\end{tabularx}
\end{table}

\textbf{CAR T-cell activation and phenotypic switching.} Parameters governing CAR T-cell activation and fate switching cannot be extracted directly from FELIX. Assuming naïve and central memory CAR T-cells activate within two days when blast numbers exceed $\sim 10^9$ cells (1/1000 of total)~\cite{kaech2002effector} and that 90\% of cells are activated over that window, we obtain $k_3\approx 10^{-9}\;\text{day}^{-1}\,\text{cell}^{-1}$, with a wide plausible range $(10^{-10},10^{-8})$. The activated-cell division rate $k_4=0.1\;\text{day}^{-1}$ is taken to be of the same order as, but slower than, the effector division rate. The phenotypic-switching threshold is set to $B_{1/2}=10^{9}$ cells with range $(10^8,10^{10})$; above this burden, activated cells differentiate predominantly into effectors, and below it into memory cells. As noted in \Cref{modeldev}, $B_{1/2}$ is a phenomenological parameter rather than a directly measured cellular threshold.

\textbf{Initial conditions.} A 100\% disease burden is taken to correspond to $10^{12}$ blast cells~\cite{saygin2022measurable, roddie2024obecabtagene}, in line with our chosen carrying capacity. For the baseline simulations we set $B(0)=2\times 10^{11}$ (20\% of $K$), reflecting the typical FELIX disease-burden scale rather than maximum burden. FELIX administered the CAR T-cell product in two doses nine days apart, totalling $C_0=4.1\times 10^{8}$ cells; for simplicity we treat the dose as a single injection at $t=0$. The infused product is partitioned by phenotype following Figure S3 of~\cite{roddie2024obecabtagene}: na\"ive and central memory cells (46.8\%) are assigned to $M$, while effector memory plus terminally differentiated cells (53.2\%) are distributed equally across the six effector division compartments to represent the mixture of proliferative potentials. This gives $M(0)=0.468\,C_0$, $A(0)=0$, and $E_i(0)=0.532\,C_0/N$ for $i=1,\ldots,N$.

\begin{table}[H]
\centering
\caption{Baseline initial conditions used in numerical simulations.}
\label{tab:initcond}
\begin{tabular}{@{}ccc@{}}
\toprule
\textbf{variable} & \textbf{description} & \textbf{value}\\
\midrule
$B(0)$ & initial blast burden & $2\times10^{11}$ \\
$C_0$ & total infused CAR T-cell dose & $4.1\times10^8$ \\
$M(0)$ & initial memory CAR T-cells & $0.468C_0$ \\
$A(0)$ & initial activated CAR T-cells & $0$ \\
$E_i(0)$ & initial effector CAR T-cells, & $0.532C_0/N$ \\
 & $i=1,\ldots,N$ & \\
\bottomrule
\end{tabular}
\end{table}

\subsection{Numerical methods} \label{hybridalgo}

To simulate the model equations defined in \Cref{modeldev}, we use a hybrid numerical scheme that combines a deterministic ODE solver with discrete stochastic simulations. This approach is necessary because during successful treatment, the tumour population can fall to very low levels, where stochastic effects become biologically significant and the continuous approximation breaks down. In particular, tumour elimination, while biologically meaningful, cannot be achieved using a purely deterministic formalism (see dynamical system analysis of \Cref{bifurc}). Dynamics involving small populations must therefore be treated discretely. At the same time, many CAR T-cell populations remain sufficiently large that their high-frequency reactions can be accurately and more efficiently approximated by deterministic dynamics. We therefore use a reaction-based hybrid method, in which individual reactions, rather than entire species, are classified as stochastic or deterministic. This is important because a given species may simultaneously participate in both stochastic and deterministic reactions.

In the early stages of treatment, all populations are large and the system is well-approximated by deterministic dynamics. We solve \Cref{eq_blast_1}-(\ref{memory1}) using MATLAB's \texttt{ode45} solver, which is based on an explicit Runge--Kutta (4,5) formula, the Dormand-Prince pair. When specific conditions are met, we switch to our hybrid algorithm that combines discrete event simulation with the \texttt{ode45} numerical solver based on the algorithm developed and analysed in \cite{sto_sim, base_sto_algo, discuss_sto_algo}.

At each macro-step, a reaction is treated deterministically when both its expected next-event time is shorter than $\delta t = 0.1\;\text{day}$ and its reactant population exceeds $\Lambda = 10^{3}$; otherwise it is sampled stochastically. The reaction network, the full hybrid update scheme, and the supporting derivations are given in \Cref{appA}.

\subsection{Dynamical system analysis} \label{bifurc}

To gain qualitative insight into the model behaviour and long-term treatment outcomes, we conduct a dynamical systems analysis of the BEAM model. Since exact solutions are analytically intractable, we analyse the system by identifying steady states and classifying their stability. The full dimensionless system, its steady-state algebra, and the linear stability analysis are given in \Cref{appB}; we summarise the results here.

The system has four steady states. Two are trivial: $P_0=(B^*=0,\mathbf{0})$ (tumour elimination, with full CAR T-cell contraction) and $P_1=(B^*=K,\mathbf{0})$ (tumour escape to carrying capacity, with all CAR T-cells dead). The other two, $P_\pm$, are coexistence states with non-zero blast and CAR T-cell populations. Linear stability analysis shows that $P_0$ is a saddle: small deviations from zero blasts grow, so tumour elimination is not a stable deterministic outcome, which motivates the hybrid stochastic--deterministic algorithm of \Cref{hybridalgo}. The escape state $P_1$ is linearly stable for the biologically relevant regime $0<B_{1/2}<K$.

\begin{figure}[h!]
    \centering
    \includegraphics[width=0.75\linewidth]{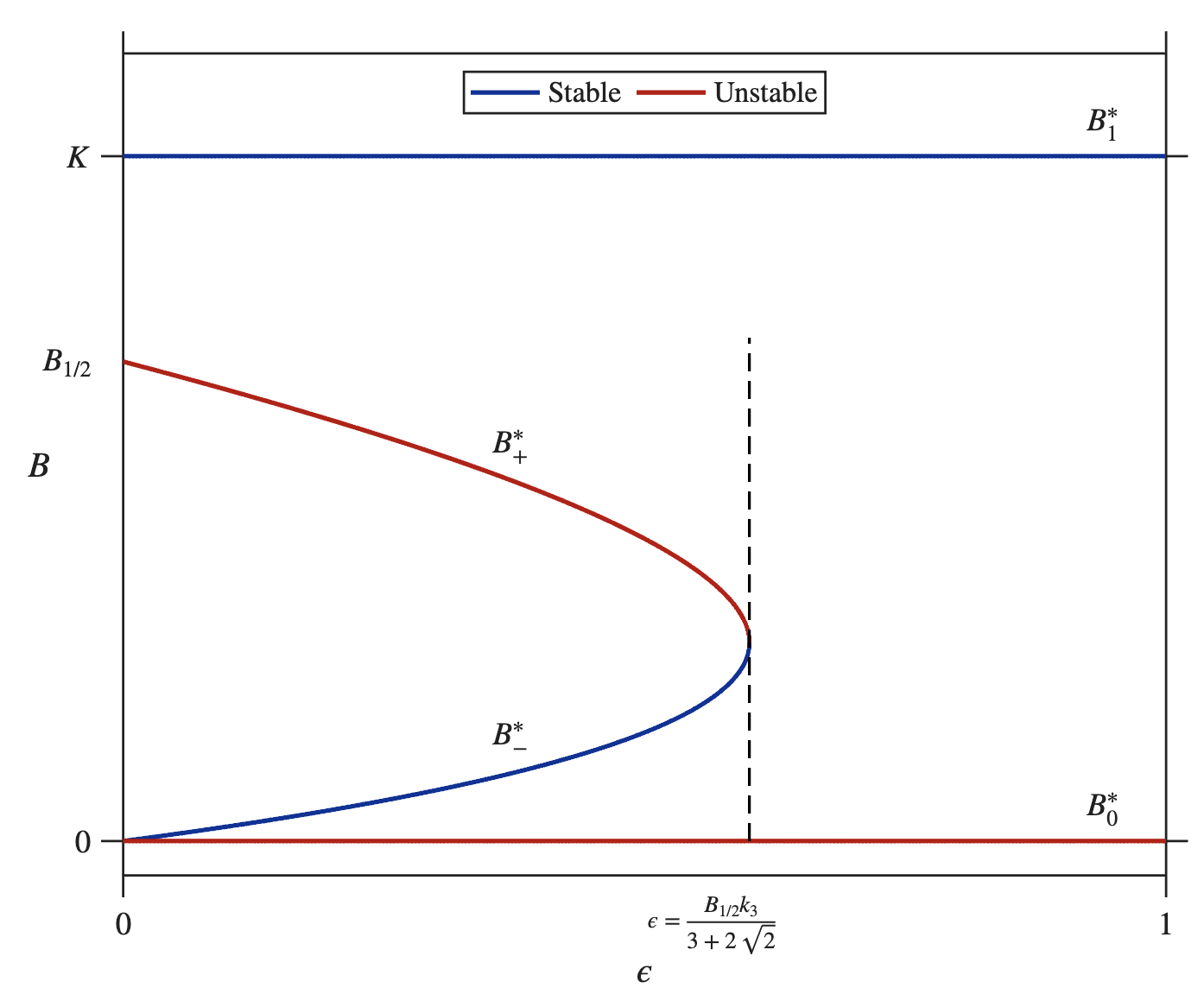}
    \caption{Bifurcation diagram of the steady-state tumour burden versus the memory CAR T-cell death rate $\epsilon$. A fold bifurcation occurs where $B^*=B_{1/2}(\sqrt{2}-1)$. Stable steady states are in blue, unstable are in red. For clarity a larger $B_{1/2}$ is used than the baseline estimate. For $B_{1/2}=10^9$, $K=10^{12}$ the coexistence burden lies close to the horizontal axis.}
    \label{fig:c2birfurc}
\end{figure}

The coexistence pair exists only below a fold bifurcation in the memory CAR T-cell death rate,
\begin{equation}\label{foldbif}
\epsilon < \frac{k_3 B_{1/2}}{3+2\sqrt{2}}.
\end{equation}
After this bifurcation, $P_+$ is unstable and $P_-$, corresponding to tumour dormancy, is stable for the parameter regimes considered. Above the threshold for this bifurcation, only escape or elimination are feasible; below it all three outcomes are possible, depending on initial conditions. Critically, the dimensional dormant blast burden
\begin{equation}\label{dimensional_coexistence_main}
B^*_{-}=\frac{k_3B_{1/2}-\epsilon - \sqrt{(k_3B_{1/2}-\epsilon)^2 - 4 k_3 B_{1/2}\epsilon}}{2k_3}
\end{equation}
is determined entirely by the memory-related parameters $k_3$, $B_{1/2}$ and $\epsilon$, independent of the effector parameters $\gamma$, $\delta$ (\Cref{appB}). The bifurcation diagram is shown in \Cref{fig:c2birfurc}.

As $\epsilon$ decreases below the threshold of \eqref{foldbif}, a bistable regime emerges in which both escape and dormancy are possible. Increasing memory lifespan further (decreasing $\epsilon$) lowers the dormant blast burden $B^*_-$ towards zero. Depending on other parameter values, the tumour burden may oscillate around this equilibrium. Notably, dormancy is controlled exclusively by the memory-related parameters $(k_3, B_{1/2}, \epsilon)$: effector CAR T parameters change the CAR T-cell levels required to maintain dormancy but not the dormant blast burden itself.

\section{Results}\label{results2}

\subsection{Initial response and stochastic elimination}

\begin{figure}[t]
    \centering
        \includegraphics[width=1\linewidth]{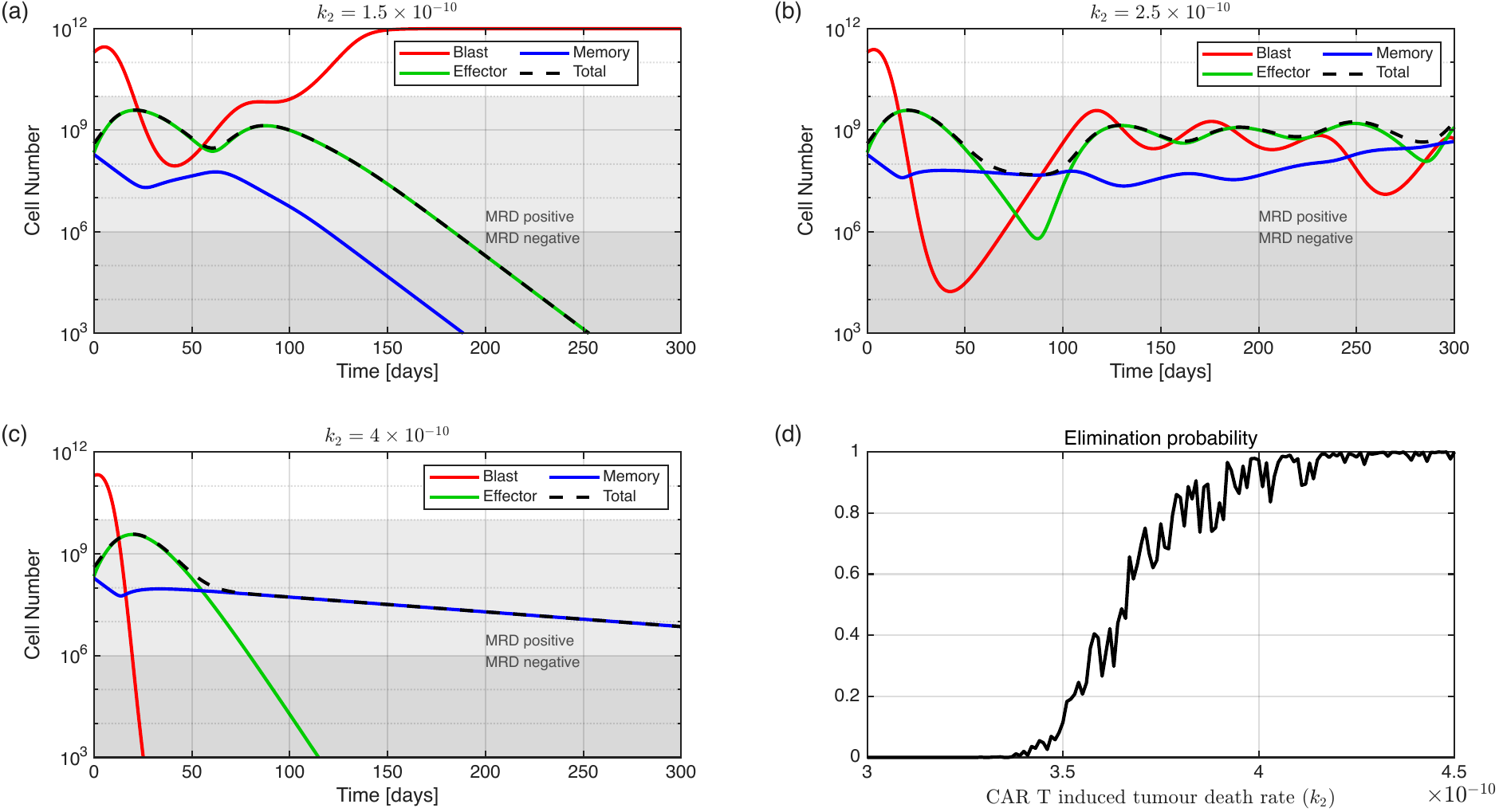}
        \caption{Numerical simulations of Equations (\ref{eq_blast_1})--(\ref{memory1}) for baseline parameters (\Cref{tab:ndparamc2}, \Cref{tab:initcond}) and varying killing rate $k_2$, over 300 days. Blast cells $B$ (red), effector CAR T-cells $\sum_i E_i$ (green), memory-like cells $M+A$ (blue), and total CAR T-cells (dashed black). The $10^6$-blast threshold separates MRD-positive and MRD-negative disease (shaded). (a) Low $k_2=1.5\times10^{-10}$: the tumour stays above MRD and escapes. (b) Intermediate $k_2=2.5\times10^{-10}$: burden falls transiently below MRD, then regrows (oscillatory relapse or dormancy-like dynamics). (c) High $k_2=4\times10^{-10}$: burden falls below MRD and is stochastically eliminated. (d) Probability of stochastic elimination versus $k_2$ ($10^3$ realisations).}
        \label{fig:treatment_success}
\end{figure}

Numerical simulations of the coupled ODE system~\eqref{eq_blast_1}--\eqref{memory1} are carried out in MATLAB using the hybrid algorithm of \Cref{hybridalgo}. Figure~\ref{fig:treatment_success} illustrates how the system's qualitative behaviour changes as the CAR T-cell induced tumour death parameter $k_2$ is varied. All other parameters are held fixed at their baseline values listed in \Cref{tab:ndparamc2} and the baseline initial conditions in \Cref{tab:initcond}. Biologically, increasing $k_2$ corresponds to increasing the effective cytotoxic activity of the CAR T-cells. Since $k_2$ is a lumped killing parameter, this may reflect changes in tumour--CAR T encounter, antigen recognition or per-cell killing efficiency. 

Following CAR T-cell injection, the effector population expands rapidly, reaching a peak approximately 3 weeks post-infusion, before contracting. During this period, effector CAR T-cells target and kill blast cells, leading to a sharp decline in the blast population. We distinguish three related outcomes. (i) A successful initial response is recorded if the blast count crosses the minimal residual disease (MRD) detection threshold of $10^6$ cells~\cite{saygin2022measurable} at any point during therapy. Conditional on an initial response, two long-term outcomes are possible: (ii) stochastic elimination, in which the discrete blast count subsequently reaches zero, and (iii) relapse, in which residual blasts regrow toward carrying capacity. Crossing the MRD threshold is therefore a necessary but not sufficient condition for durable treatment success.

In Figure~\ref{fig:treatment_success}a, the blast count remains within MRD levels throughout, and the tumour eventually regrows toward its carrying capacity $K$, indicating treatment failure. As $k_2$ increases, CAR T-cells become more effective and we distinguish two types of initial response. In Figure~\ref{fig:treatment_success}b, the blast count dips below MRD but does not reach zero, eventually beginning to rise again. In contrast, Figure~\ref{fig:treatment_success}c shows full tumour eradication ($\min_t B(t) = 0$). In the latter case, effector CAR T-cells contract following tumour elimination and, by three months, the population consists predominantly of memory CAR T-cells. The post-contraction CAR T-cell count is approximately two orders of magnitude lower than its peak, consistent with clinical observations from the FELIX study (see Figure S13 of~\cite{roddie2024obecabtagene}).

For the baseline parameters, the deterministic system has a stable dormant coexistence state because $\epsilon < k_3B_{1/2}/(3+2\sqrt{2})$ (\Cref{foldbif}; \Cref{appB}). Changing $k_2$ does not change the existence or the blast burden of this dormant equilibrium; rather, it rescales the dimensional CAR T-cell abundance required for coexistence and changes the trajectory followed after infusion. In particular, increasing $k_2$ strengthens the early cytotoxic response, changes whether the trajectory enters the basin of attraction of tumour escape or dormancy, and increases the probability that the blast population is driven to sufficiently low numbers for stochastic extinction.

Because the hybrid algorithm treats reactions discretely once populations become small (\Cref{hybridalgo}), residual blasts can undergo exact stochastic extinction. Figure~\ref{fig:treatment_success}d shows the probability of eliminating all cancer cells as a function of $k_2$, computed over $10^3$ realisations: across a wide range of $k_2$, both elimination and relapse are possible outcomes.

\textbf{Drivers of initial response.} Since the initial response is characterised by the minimum blast cell count achieved during therapy, it is primarily governed by the parameters influencing the early-phase tumour response. When the blast population is well below carrying capacity ($B\ll K$), the governing ODE simplifies to:
\begin{equation}
\frac{\mathrm{d}B}{\mathrm{d}t} \approx B\left(k_1 - k_2\sum_i E_i\right),
\end{equation}
which describes exponential growth or decay depending on the relative balance between the tumour growth rate $k_1$ and the cumulative tumour-killing activity of effector CAR T-cells, $k_2\sum_i E_i$. Initial response outcomes therefore depend on three key factors: (a) tumour growth rate ($k_1$), (b) CAR T-cell killing efficacy ($k_2$), and (c) the total number of effector CAR T-cells ($\sum_i E_i$). The integrated effector exposure $\int_0^t \sum_i E_i(s)\,ds$ is in turn shaped by the infused dose $C_0$, the initial memory-versus-effector composition of the product, and the effector-chain dynamics governed by $\gamma$ (division rate) and $\delta$ (terminal loss rate). Increasing $\delta$ consistently impairs the initial CAR T response. The dependence on the effector division rate $\gamma$ is non-monotonic: too little proliferation limits effector expansion, while excessive proliferation drives faster transit through the finite division chain and therefore earlier terminal loss; hence, the model predicts an intermediate optimum for tumour control. In the high-burden regime ($B\gg B_{1/2}$), the minimum blast count is less sensitive to the initial tumour burden $B_0$ than to parameters controlling effector exposure. This is a direct consequence of two modelling choices: the activation-to-effector fate switch saturates when $B\gg B_{1/2}$, and individual CAR T-cells are not assigned a finite killing capacity. Clinical evidence associating $B_0$ with response depth and relapse risk indicates that this is an oversimplification; functional exhaustion of effector CAR T-cells under prolonged antigen engagement~\cite{lai2025impact, dolina2021cd8} is one mechanism not captured by our finite division chain, and we return to this limitation in \Cref{discussion}.

\begin{figure}[t]
    \centering
    \includegraphics[width=1\linewidth]{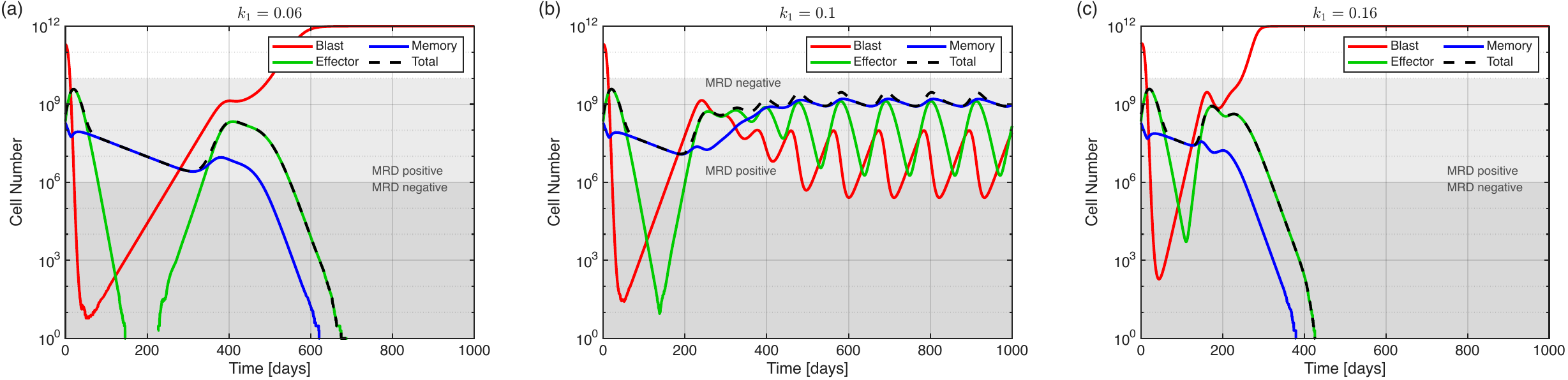}
    \caption{Numerical simulations of the model equations for varying tumour growth rate $k_1$, over 3 years, with $\epsilon = 0.017\;\mathrm{day}^{-1}$ (faster than the baseline $0.01$, so late relapse completes within the window) and all other parameters at baseline (\Cref{tab:ndparamc2}, \Cref{tab:initcond}). (a) Slow growth ($k_1 = 0.06\;\mathrm{day}^{-1}$): delayed escape after a deep initial response. (b) Dormancy ($k_1 = 0.1\;\mathrm{day}^{-1}$). (c) Fast growth ($k_1 = 0.16\;\mathrm{day}^{-1}$): escape after progressively weakening secondary responses. Colours as in \Cref{fig:treatment_success}.}
    \label{fig:k1scenarios}
\end{figure}

\subsection{Relapse}
We now focus on treatment scenarios where the tumour burden falls below MRD detection thresholds but the blasts are not fully eliminated. In such cases, the tumour may eventually regrow, resulting in relapse or long-term control that depends on the interplay between tumour proliferation, immune persistence, and effector cell dynamics. \Cref{fig:k1scenarios} illustrates three qualitatively distinct outcomes, obtained by varying the tumour growth rate $k_1$ while holding other parameters fixed. In this section, \emph{relapse} refers to regrowth of residual blasts after an initial MRD response, whereas ``dormancy'' refers to long-term immune-mediated control without stochastic elimination.

\textbf{Relapse in highly proliferative tumours.} For large tumour growth rates ($k_1 = 0.16\;\mathrm{day}^{-1}$ in \Cref{fig:k1scenarios}c), the tumour burden begins to rise rapidly once the effector CAR T-cell population contracts. In response, memory CAR T-cells differentiate into new effectors, temporarily suppressing the tumour. Each such secondary wave depletes the memory pool further without restoring it, since memory is replenished only when tumour burden is low enough to favour the memory lineage. Successive secondary responses therefore weaken until the tumour escapes towards its carrying capacity.

\textbf{Long-term tumour control (dormancy).} At intermediate tumour proliferation rates ($k_1 = 0.1\;\mathrm{day}^{-1}$ in \Cref{fig:k1scenarios}b), the immune system can maintain tumour control through a dynamic equilibrium. Fewer effector cells are needed to contain the less aggressive tumour, which allows the memory population to remain sufficiently robust to continually replenish effectors. This leads to a persistent state of tumour dormancy, in which both blast and CAR T-cell populations stabilise. The full mathematical characterisation of this equilibrium and its linear stability is provided in \Cref{bifurc} and \Cref{appB}. When in an equilibrium state, the tumour burden may be constant or oscillate around an equilibrium value.

\textbf{Relapse after deep initial response.} For low tumour growth rates ($k_1 = 0.06\;\mathrm{day}^{-1}$ in \Cref{fig:k1scenarios}a), the initial response is particularly strong, reducing the blast count to very low levels. When tumour numbers are this low, the likelihood of encounters between tumour and CAR T-cells drops significantly, allowing remaining cancer cells to evade immune detection and slowly regrow. The tumour burden remains below MRD thresholds for several months, during which the memory CAR T-cell population progressively depletes. The tumour eventually rebounds at a point when the memory CAR T-cell population is insufficient to mount a successful secondary response, leading to escape of the tumour.

Varying $k_1$ does not change the dormant tumour burden $B_-^*$, which is determined by $k_3$, $B_{1/2}$ and $\epsilon$, but it does change the CAR T-cell levels. In particular, the dimensional effector population required to balance tumour growth scales approximately with $k_1/k_2$, since tumour control requires $k_2\sum_iE_i \approx k_1(1-B/K)$. Thus, faster-growing tumours require larger effector populations for control and can deplete the memory pool through repeated activation, whereas very slow-growing tumours may remain at low abundance for long periods, allowing memory CAR T-cells to decay before the residual tumour becomes large enough to trigger a strong secondary response.

This produces a non-monotone dependence of relapse outcome on tumour growth rate. Intermediate growth rates can generate enough antigenic stimulation to maintain the memory--effector feedback loop. In contrast, fast tumours outpace or exhaust this feedback, and very slow tumours can remain hidden below effective immune-stimulation levels until CAR T-cell persistence has waned. 

Because slow-growing tumours can be driven to very low cell numbers during the initial response, stochastic elimination may occur before delayed relapse. We quantify the dependence of treatment outcome on tumour growth rate through a parameter sweep over $k_1$ in \Cref{fig:bifurc_persistence}, using the same hybrid algorithm and classification criteria as in \Cref{fig:treatment_success}d.

\subsection{Increasing immune persistence}

Since lack of CAR T-cell persistence is clinically associated with relapse \cite{roddie2024obecabtagene, park2018long, stein2019tisagenlecleucel, melenhorst2022decade}, we now ask how persistence can be increased. In our model, memory CAR T-cells are divided into two compartments: resting memory cells ($M$) and activated memory cells ($A$). We explore two strategies to increase immune persistence: increasing memory cell lifespan (by decreasing $\epsilon$), and biasing CAR T-cell fate toward memory rather than effector phenotypes (by increasing $B_{1/2}$).

\begin{figure}[t]
    \centering
    \includegraphics[width=1\linewidth]{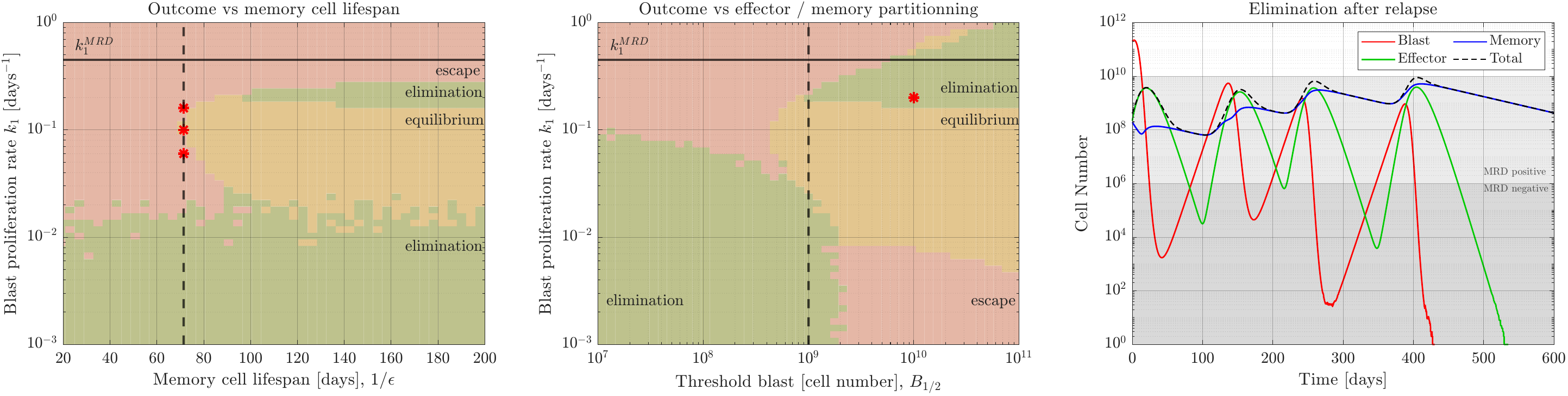}
    \caption{Treatment outcome as immune persistence is increased, over $T=365$ days; escape red, dormancy/equilibrium orange, elimination green. Computed using 500 hybrid realisations at each grid point and colouring it by the most frequent outcome, classified as elimination if the blast count reaches zero, escape if it exceeds $9\times10^{11}$ cells or all immune cells are lost, and dormancy (equilibrium) otherwise. (a) Outcome versus memory CAR T-cell lifespan $1/\epsilon$ and growth rate $k_1$ on a $50\times50$ grid ($k_1\in[10^{-3},10^{0}]\;\mathrm{day}^{-1}$, $1/\epsilon\in[20,200]$ days); baseline for other parameters (\Cref{tab:ndparamc2}, \Cref{tab:initcond}). Red stars mark the \Cref{fig:k1scenarios} simulations. (b) Outcome versus differentiation threshold $B_{1/2}$ and $k_1$ on a $50\times50$ grid ($B_{1/2}\in[10^{7},10^{11}]$ cells; dashed line at baseline $B_{1/2}=10^{9}$). (c) A representative elimination-after-relapse trajectory for the starred point in (b), with $B_{1/2}=10^{10}$ and other parameters at baseline.} 
    \label{fig:bifurc_persistence}
\end{figure}

\textbf{Increasing memory CAR T-cell lifespan.} Figure \ref{fig:bifurc_persistence}(a) shows how the treatment outcome varies with tumour proliferation rate $k_1$ and memory cell lifespan $1/\epsilon$. Two thresholds are identified:
\begin{itemize}
    \item $k_1^{\text{MRD}}$: the maximum $k_1$ for which tumour burden falls below minimal residual disease (MRD) levels at any point during treatment and,
    \item $k_1^*$: the maximum tumour growth rate below which elimination can be achieved during the initial response. Since elimination is stochastic, we choose $k_1^*$ to be the proliferation rate below which stochastic elimination is the most frequent outcome among the 500 simulations.
\end{itemize}
The initial depth of tumour reduction is governed by effector CAR T-cell cytotoxicity and expansion, not memory dynamics. Hence, $k_1^{\text{MRD}}$ is independent of $\epsilon$, and $k_1^*$ is only weakly affected by $\epsilon$ over the range shown. However, for growth rates between these two thresholds, the outcome depends on memory cell persistence. Short memory cell lifespans result in tumour escape (red region) while longer lifespans can sustain immune surveillance and maintain tumour dormancy (orange). Increasing memory cell lifespan broadens the range of tumour growth rates over which control is maintained, even if complete elimination is not achieved. This supports the therapeutic value of engineering CAR T-cells for long-term persistence. The three red stars on the dashed line in \Cref{fig:bifurc_persistence}(a) mark the simulations of \Cref{fig:k1scenarios}, illustrating the three outcomes (escape, dormancy and delayed escape) along a single vertical slice at $\epsilon = 0.017\;\mathrm{day}^{-1}$.

\textbf{Modulating CAR T-cell fate: the role of $B_{1/2}$.} In our model, the parameter $B_{1/2}$ determines the tumour burden threshold above which activated CAR T-cells differentiate mostly into effectors; below this, they revert to memory phenotypes. Figure \ref{fig:bifurc_persistence}(b) shows that low $B_{1/2}$ values bias cells toward the effector fate, enabling strong initial tumour killing and increasing the range of $k_1$ values for which elimination occurs. High $B_{1/2}$ values promote memory differentiation, reducing immediate cytotoxicity but improving long-term persistence and increasing the dormancy regime.

A trade-off emerges: while high effector differentiation (low $B_{1/2}$) favours initial elimination for slowly proliferating tumours, it may limit the system's ability to respond to recurrence. Conversely, higher $B_{1/2}$ supports immune surveillance for higher tumour proliferation at the cost of early cytotoxic strength. This supports the hypothesis that CAR T-cell product characteristics should be tailored to individual patients, as varying intrinsic tumour proliferation rates may require distinct phenotypic profiles to achieve optimal therapeutic outcomes. 

\textbf{Elimination after relapse.} For a small range of parameters, oscillatory tumour dormancy can transition into full elimination (green regions in Figures \ref{fig:bifurc_persistence}, particularly the upper-right corner of (b)). A representative simulation is shown in Figure \ref{fig:bifurc_persistence}(c), with $B_{1/2}=10^{10}$ cells (above baseline) and all other parameters at baseline values. As the tumour remains in a dormant state with sustained immune surveillance, the interaction between tumour regrowth and delayed effector expansion generates large-amplitude oscillations in tumour burden over several cycles. Combined with high immune persistence, each successive blast peak is met by a strong effector response, and stochastic extinction occurs after several such cycles rather than during the initial response.

However, this outcome requires a larger CAR T-cell expansion during the relapse phase than during the initial response. Clinical data from Figure S13 in \cite{roddie2024obecabtagene} suggests that such a robust secondary expansion may not occur naturally. This raises the possibility that a delayed second infusion of CAR T-cells could drive elimination, provided that a sufficient pool of memory CAR T-cells remains. While this strategy has not yet been extensively tested in clinical trials, largely due to concerns about toxicity \cite{hardiansyah2019quantitative}, it may warrant future exploration for patients at risk of relapse.

\subsection{Loss of CD19 antigen.} \label{antigenescape}

\begin{figure}[t]
    \centering
    \includegraphics[width=.8\linewidth]{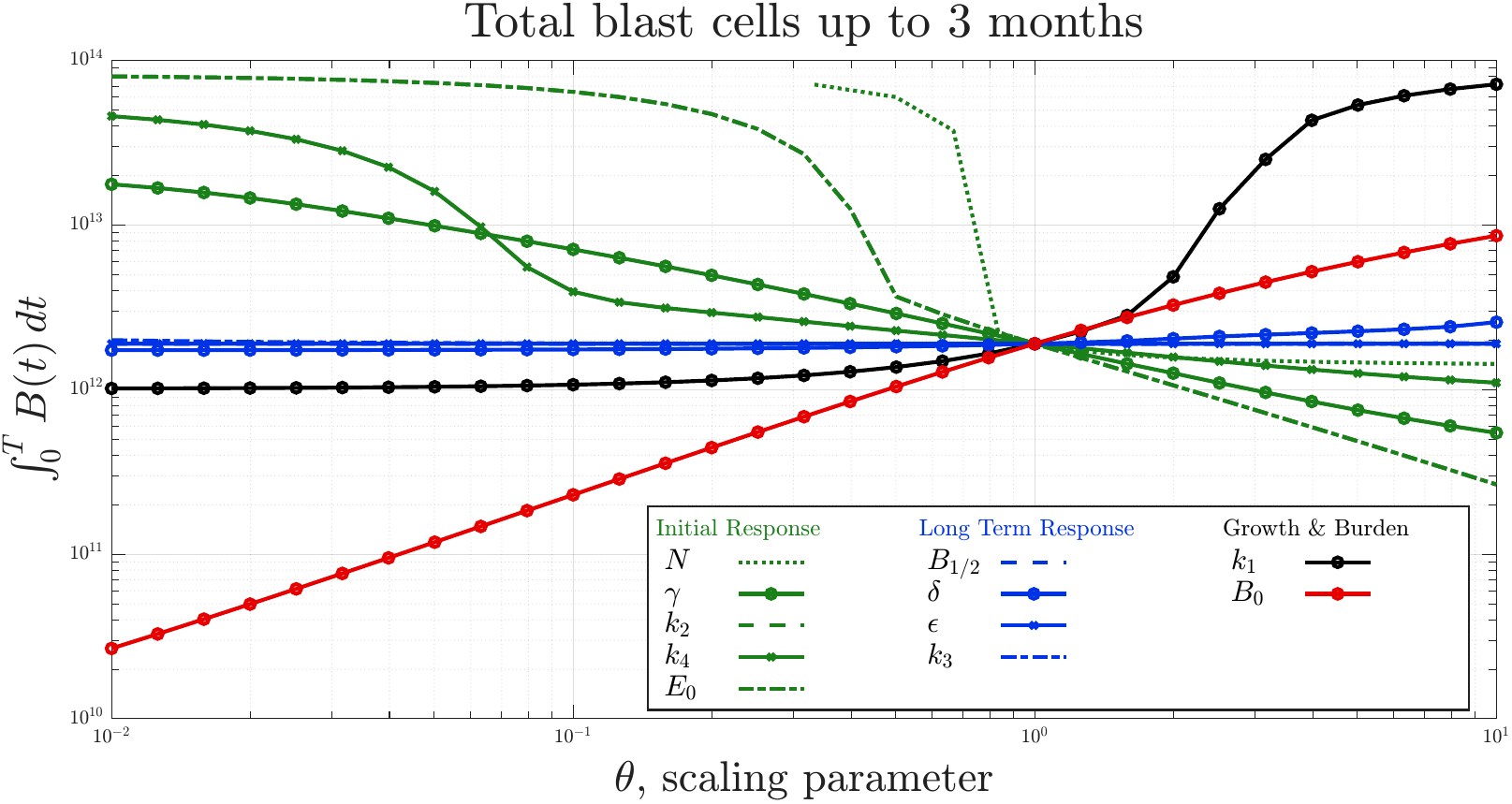}
    \caption{Relapse via CD19-negative antigen escape. Cumulative CD19-positive blast burden $\int_0^T B(\tau)\,d\tau$ up to $T=3$ months (units cell-days), under one-at-a-time scaling of each parameter by a factor $\theta$ ($\theta=1$ is baseline; \Cref{tab:ndparamc2}, \Cref{tab:initcond}); for $E_0$ the total initial effector-like population is scaled with the equal $E_i$ split retained. Parameters are grouped by effect on the integral: ``Initial Response'' ($N$, $\gamma$, $k_2$, $k_4$, $E_0$) shape the early effector wave; ``Long Term Response'' ($B_{1/2}$, $\delta$, $\epsilon$, $k_3$) act on slower memory dynamics and barely affect the 3-month integral; ``Growth \& Burden'' ($k_1$, $B_0$) set initial accumulation. Lower values mean fewer opportunities for antigen-negative escape.}
    \label{fig:loss_cd19}
\end{figure}

In the FELIX study, among 31 patients whose expression of CD19 in blasts was assessed at the time of relapse, 15 had a relapse with CD19-negative blasts (Supplementary Table S15 of \cite{roddie2024obecabtagene}). To date, validated predictors for antigen negative relapse remain limited, highlighting the need for modelling work. We study CD19-negative antigen escape in our model by introducing a rate $\kappa$ at which a CD19 expressing blast cell (CD19-positive) loses its antigen during cellular division.

Since we expect initial numbers of CD19$^-$ cells to be small, stochastic effects must be captured to accurately model relapse. To this end, we estimate the probability of relapse due to loss of antigen. Let $X(t)$ be the cumulative number of leukaemic blast cells that have lost the CD19 antigen during cellular division in the time interval $[0,t]$. Assuming a constant rate of antigen loss, $X(t)$ is a Poisson random variable that follows $X(t) \sim  \text{Poisson}(\lambda(t))$ where 
\begin{equation} \label{cd19_poisson_mean}
    \lambda(t) = \kappa k_1 \int_0^t B(u)\text{d}u. 
\end{equation}
This expression follows from the stochastic reaction network in \Cref{tab:reaction_network}, where logistic tumour growth is represented by separate blast birth and density-dependent death reactions. Antigen loss is assumed to occur during blast division, and CD19-positive blast divisions therefore occur at rate $k_1B(t)$. If logistic growth were instead implemented as a single net birth process with rate $k_1B(1-B/K)$, then the integrand in \eqref{cd19_poisson_mean} would need to be replaced by $B(t)(1-B(t)/K)$.

This implies that, on average, $\lambda(t)$ CD19 negative cells are created in $[0,t]$ ($\mathbb{E}[X(t)] = \lambda (t)$). Since we are interested in the probability of CD19 negative escape, we further model the birth-death process of these cells. We assume that when a CD19 negative cell divides (at rate $b$), both daughter cells lack the CD19 antigen and we let $d$ be their death rate with $b > d$. If we consider a single cell, the probability that the lineage originating from this single cell eventually goes extinct (i.e., is absorbed at $0$) is given by the classical result
\begin{equation}
    \mathbb{P}(\text{extinction}) = \frac{d}{b} =: \beta. 
\end{equation}
Now suppose there are initially $k$ such independent CD19 negative cells. Then, under the assumption of independent lineages, the probability that all $k$ lineages go extinct is:
\begin{equation}
    \mathbb{P}(\text{all extinct} | X = k) = \beta^k
\end{equation}
Since the number of CD19 negative blasts is itself a random variable, the overall probability that all CD19 negative lineages go extinct is then given by the law of total probability
\begin{subequations} 
\begin{align}
    \mathbb{P}(\text{all extinct}) &= \sum_{k = 0}^{\infty}  \mathbb{P}(\text{all extinct} | X = k) \times \mathbb{P}(X = k)\\
    &= \sum_{k = 0}^{\infty} \beta^k \times \mathbb{P}(X = k) = \mathbb{E}[\beta^X].
\end{align}
\end{subequations}
Since $X \sim \text{Poisson}(\lambda)$, its moment-generating function is
\begin{equation}
    \mathbb{E}[e^{sX}] = e^{\lambda \left(e^s-1\right)}. 
\end{equation}
Therefore
\begin{equation}
     \mathbb{P}(\text{all extinct}) = \mathbb{E}[\beta^X] = \mathbb{E}[e^{X \log(\beta)}] = e^{ \lambda \left(\beta-1\right)}
\end{equation}
and finally, we obtain 
\begin{equation} \label{prob_cd19}
     \mathbb{P}(\text{CD19$^-$ escape}) = 1 -  \exp{  \left[ - \left(  \kappa k_1 \int_0^t B(s) ds\right)  \left(  1 - \frac{d}{b}\right) \right]}.
\end{equation}
We explore how the cumulative CD19$^+$ blast burden, given by the integral $\int_0^t B(\tau) d\tau$ in \Cref{prob_cd19}, changes with key model parameters. This quantity governs the number of opportunities for antigen loss during cell division, and therefore directly influences the risk of CD19$^-$ escape.

The dependence of this cumulative burden on the initial CAR T-cell response can also be understood analytically. During the early high-burden phase, when $B\gg B_{1/2}$, the activated-cell population rapidly approaches an exponentially decaying form $A(t)\simeq A_*e^{-k_4t}$, where $A_*$ is the effective activated-cell amplitude following the short initial activation transient. The effector division chain is then linear and can be solved explicitly for $E(t)=\sum_{i=1}^N E_i(t)$. Treating this early-time effector response as prescribed, the blast equation becomes a logistic equation with time-dependent growth rate $k_1-k_2E(t)$, with solution
\begin{equation}
    B(t)=\frac{B_0e^{\Psi(t)}}{1+\frac{k_1B_0}{K}\int_0^t e^{\Psi(s)}\,\text{d}s}, \qquad \Psi(t)=k_1t-k_2\int_0^t E(s)\,\text{d}s.
\end{equation}
Consequently, the cumulative blast burden considered below is
\begin{equation}
    \int_0^T B(t)\,dt=\frac{K}{k_1}\log\left(1+\frac{k_1B_0}{K}\int_0^T e^{\Psi(s)}\,ds\right).
\end{equation}
The full derivation and the corresponding explicit solution of the effector division chain are given in \cref{appC}. This approximation provides an analytical interpretation of the parameter dependencies observed in \Cref{fig:loss_cd19}.

\Cref{fig:loss_cd19} shows that parameters influencing the early effector CAR T-cell response have a significant effect on this integral. Increasing the killing rate $k_2$, the division rates $k_4$ and $\gamma$, the number of effector division stages $N$, or the initial effector-like CAR T-cell count $E_0=\sum_iE_i(0)$ (with $M(0)$ and $A(0)$ kept at their baseline values) can reduce the total number of tumour cells by almost an order of magnitude. In contrast, parameters primarily associated with long-term immune-persistence ($\epsilon$, $\delta$, $k_3$, $B_{1/2}$) have a weaker effect on the 3-month integral, because the memory pool only becomes the dominant source of effector replenishment after the initial effector wave has contracted. Notably, the initial tumour burden $B_0$ exerts a strong influence on escape probability. This is because, in both elimination and dormancy scenarios, the majority of tumour cell divisions, and thus opportunities for antigen loss, occur early in treatment.

Two parameters dominate the integral and therefore the escape probability: $k_1$ and $B_0$. Of these, $k_1$ is a property of the patient's disease and is not directly modifiable, whereas $B_0$ can be reduced before infusion through cytoreductive or bridging therapy~\cite{feuchtinger2024approaches, shahid2021bridging}. \Cref{prob_cd19} shows that, for a fixed antigen-loss probability per division, lowering the cumulative CD19-positive blast burden before immune control is established directly reduces the probability that a surviving CD19-negative lineage is generated. Combined with the effector-response parameters in the green group, this provides a quantitative argument for the clinical practice of disease reduction prior to CAR T-cell infusion in patients at risk of antigen-negative relapse.

\subsection{Blasts hidden in an immune-privileged niche}

\begin{figure}[t]
    \centering \includegraphics[width=1\linewidth]{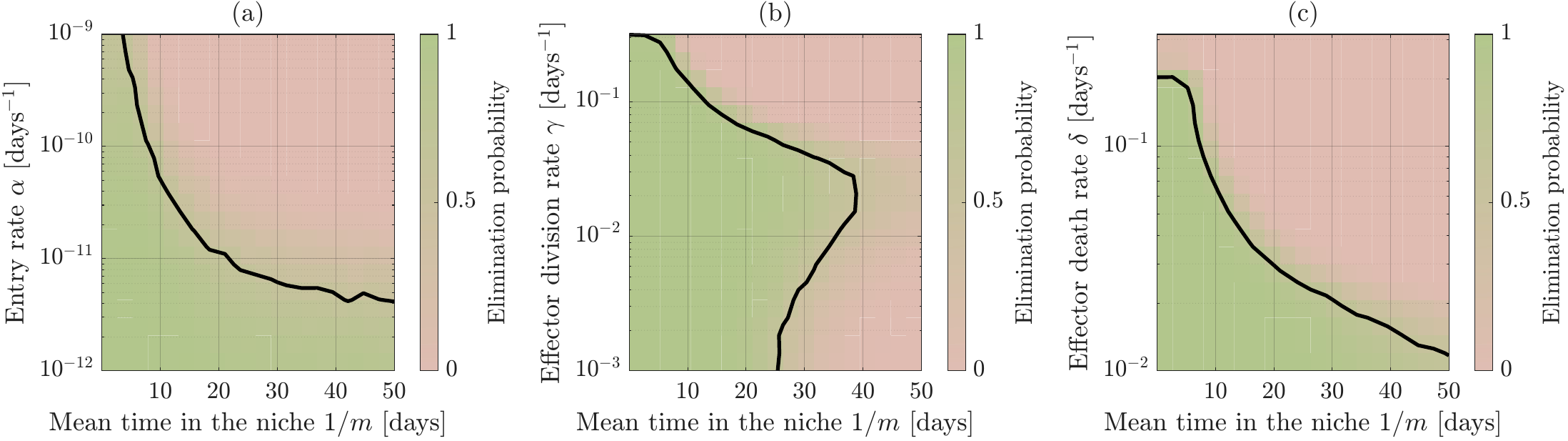}
    \caption{Treatment outcomes when blasts can hide in a niche, at 6 months (500 simulations per point on a $20\times20$ grid; black line marks elimination probability $0.5$). Elimination requires $B(T)=H(T)=0$, where $H$ is the hidden-niche population. Baseline parameters (\Cref{tab:ndparamc2}, \Cref{tab:initcond}) with $k_2=4\times10^{-10}\,\mathrm{day}^{-1}\mathrm{cell}^{-1}$ (so circulating blasts are eliminated without a niche) and $H(0)=0$. (a) Versus niche entry rate $\alpha\in[10^{-12},10^{-9}]\,\mathrm{day}^{-1}$ and mean residence time $1/m\in[1,50]$ days. (b) Versus effector division rate $\gamma\in[10^{-3},0.3]\,\mathrm{day}^{-1}$ ($\alpha=10^{-10}$). (c) Versus effector death rate $\delta\in[10^{-2},0.3]\,\mathrm{day}^{-1}$ ($\alpha=10^{-10}$).}
    \label{fig:niche}
\end{figure}

Some B-ALL blasts may evade clearance by entering immune-privileged niches of the bone marrow, adopting the low-proliferation, dormant state characteristic of the haematopoietic stem cells that normally occupy them \cite{szade2018hematopoietic, dander2021bone, ma2020leukemia}. Such hidden cells pose a therapeutic challenge: they can re-emerge after effector CAR T-cells contract, so sustained surveillance is needed to clear blasts leaving the niche before they re-establish disease. To investigate this mechanism, we introduce an additional hidden blast compartment $H$ into the model. Blasts enter the niche via
\begin{equation}
    B \xrightarrow{\alpha B} H,
\end{equation}
and re-enter circulation via
\begin{equation}
    H \xrightarrow{mH} B.
\end{equation}
The parameter $\alpha$ describes migration of blasts from the central marrow environment into the niche, while $m$ is the exit rate from the niche into circulation, with mean residence time $1/m$. In the absence of direct measurements of niche entry rates in B-ALL, we span a wide plausible range $\alpha\in[10^{-12},10^{-9}]\,\mathrm{day}^{-1}$, chosen so that the steady-state niche population over the simulation window covers approximately $1$ to $10^{4}$ cells. Cells within the niche are assumed to undergo negligible proliferation or death, reflecting the dormancy of HSCs observed in experimental studies \cite{szade2018hematopoietic}. They are also assumed to be protected from CAR T-cell killing while hidden. Once they exit the niche, blasts are governed by our standard tumour growth equation (\ref{eq_blast_1}).

Our analysis focuses on scenarios in which all blasts outside the niche are eliminated during the initial CAR T-cell response, thereby isolating the influence of the hidden population on subsequent relapse. To enforce this, we use the reference killing rate $k_2 = 4\times 10^{-10}\,\mathrm{day}^{-1}\mathrm{cell}^{-1}$ for which the corresponding no-niche system eliminates the tumour. Any failure of elimination in \Cref{fig:niche} is therefore attributable to the hidden compartment. Initially, the number of hidden blasts increases at rate $\alpha B$ until the blasts outside the niche have been eliminated. Once this occurs, the niche population declines through the exit of cells at rate $m$. As illustrated in \Cref{fig:niche}a, the mean residence time is the principal determinant of outcome. For example, at $\alpha = 10^{-10}\,\mathrm{day}^{-1}$, elimination is the more probable outcome when the residence time is shorter than approximately five days, since blasts emerge before effector CAR T-cell contraction has occurred; conversely, when the residence time exceeds roughly ten days, relapse becomes increasingly likely because some blasts reappear after the effector population has diminished.

Panel (a) of \Cref{fig:niche} shows that increasing either the niche entry rate $\alpha$ or the mean residence time $1/m$ reduces the probability of elimination. This suggests that clinical strategies aimed at extending the persistence of effector CAR T-cells, or interventions capable of disrupting the dormancy of niche-resident blasts, may provide effective means of mitigating relapse in B-ALL patients harbouring such immune-privileged niches.

Unlike the memory-driven control emphasised earlier, relapse from immune-privileged niches depends on the sustained presence of effector CAR T-cells. Since blasts emerge from the niche at low numbers, it is immunologically advantageous to eliminate them immediately upon exit, before their population grows large enough to interact with memory cells and trigger a broader immune response. This requires effector CAR T-cells to remain at sufficient levels over an extended period, even in the absence of detectable disease. In \Cref{fig:niche}(b,c) we plot the elimination probability of the cells that have gone through a niche when varying the effector CAR T-cell division and death rate respectively. We find that increasing the persistence of effector CAR T-cells, modelled by reducing their death rate $\delta$, substantially enhances immune surveillance efficacy, particularly when blasts exit the niche at slower rates. Panel (c) shows this monotonic effect clearly: lower $\delta$ prolongs the survival of the effector lineage and expands the region of parameter space where elimination remains likely.

The dependence on the effector division rate $\gamma$, shown in \Cref{fig:niche}b, is more subtle and non-monotone. Increasing $\gamma$ has two competing effects. On the one hand, larger $\gamma$ strengthens the initial effector burst, which suppresses the circulating blast population more rapidly and reduces the number of blasts that can enter the niche. On the other hand, faster progression through the effector division chain shortens the duration of the effector response, so that fewer effectors remain when hidden blasts later exit the niche. Consequently, an intermediate range of $\gamma$ maximises elimination probability, especially for intermediate-to-long residence times $1/m$. This biphasic behaviour is not obvious a priori and highlights a trade-off between achieving strong early cytotoxicity and maintaining sufficient effector persistence for delayed surveillance. This is the only setting in this study in which the model predicts an intermediate optimum for a CAR T-cell parameter rather than a monotonic ``more is better'' relationship.

Taken together, this analysis of niches shows that the relevant form of persistence here is effector rather than memory persistence; we revisit this distinction and its implications in \Cref{discussion}.

\section{Discussion} \label{discussion}

The BEAM model couples three CAR T-cell functional states (resting memory $M$, activated $A$, and effector cells $E_i$ across $N$ division compartments) to a logistic growth equation for the blasts, enabling us to study which forms of persistence (memory, effector, or both) prevent which forms of relapse. This is motivated by the clinical association between prolonged CAR T-cell persistence and durable remission in B-ALL \cite{park2018long, stein2019tisagenlecleucel, melenhorst2022decade, roddie2024obecabtagene}, whose mechanistic basis cannot be read off the data directly. Building on existing predator--prey and CAR T-cell models \cite{stein2019tisagenlecleucel, kirouac2023deconvolution, kirouac2024making, salem2023development, minucci2024multi}, the contribution of BEAM is the combination of a memory--activation--effector structure with deterministic stability analysis and a hybrid stochastic--deterministic simulation, which lets us distinguish deterministic dormancy from stochastic elimination and compare multiple relapse routes within one framework. Parameters were informed by the FELIX trial of obecabtagene autoleucel in adult B-ALL \cite{roddie2024obecabtagene}.

Our model predicts that a successful initial response, defined as reduction of blast counts below the minimal residual disease (MRD) threshold of $10^6$ cells, depends mainly on the intrinsic tumour growth rate $k_1$ and on the differentiation state of the infused CAR T product.

Our numerical results show that when the tumour population is brought to sufficiently small levels, stochasticity becomes important: residual blasts may either be eliminated or survive at low abundance before later regrowth. This highlights the critical importance of a robust and potent early CAR T-cell response. In the high-burden regime explored here, the minimum blast count is less sensitive to initial tumour burden than to parameters controlling early effector exposure, partly because the effector-differentiation response is saturated when $B\gg B_{1/2}$ and because the model does not impose a per-cell killing limit. This prediction may therefore be sensitive to these modelling assumptions.

For patients who achieve remission without full elimination, our model distinguished two qualitatively distinct types of relapse due to lack of immune persistence. For aggressive tumours, blasts rapidly regrow after effector CAR T-cell contraction, with progressively weaker secondary immune responses as the memory pool depletes. For low tumour growth rates, a strong initial response reduces blast counts to extremely low levels, where encounters with CAR T-cells become rare. This allows residual cancer cells to slowly regrow, eventually rebounding as the memory CAR T-cell population diminishes below a critical threshold for effective secondary responses. Slow-growing or weakly antigen-stimulating residual disease may fail to maintain sufficient CAR T-cell activation, allowing relapse after a prolonged period of apparent remission.

For both aggressive and slow-growing tumours, immune persistence can prevent relapse, instead leading to a durable state of tumour dormancy where both blast and CAR T-cell populations stabilise. This dormant blast burden is set entirely by the memory parameters ($k_3$, $\epsilon$, $B_{1/2}$), whereas effector kinetics affect only the CAR T-cell levels needed to sustain it.

We investigated two strategies to build a larger and more sustained population of memory CAR T-cells, crucial for preventing relapse. Extending the lifespan of memory CAR T-cells (by decreasing their death rate, $\epsilon$) significantly broadens the range of tumour growth rates for which long-term control or even elimination can be maintained. The parameter $B_{1/2}$ revealed a crucial trade-off. While a lower $B_{1/2}$ (favouring effector differentiation) leads to stronger initial tumour killing and increases the likelihood of elimination during the initial response, it limits the capacity for long-term surveillance. Conversely, a higher $B_{1/2}$ (favouring memory differentiation) supports long-term immune surveillance at the cost of immediate cytotoxic strength. Thus, the model predicts a trade-off between immediate cytotoxicity and durable surveillance. Effector-biased products may be better suited against slower tumours but may be weaker when there is recurrence, whereas memory-biased products are able to maintain surveillance in more proliferative disease, although this rarely leads to elimination. This strongly suggests that CAR T-cell phenotype characteristics should be tailored to individual patients and their intrinsic tumour proliferation rates to achieve optimal therapeutic outcomes.

For antigen-negative escape, modelled as CD19 loss during blast division, we showed that parameters shaping the early effector response reduce the cumulative CD19$^{+}$ blast burden most effectively, whereas long-term persistence parameters contribute little over the three-month window considered. Because most divisions occur before control is established, initial tumour burden $B_0$ is a primary driver of antigen-negative relapse which emphasises the importance of reducing disease burden before CAR T-cell infusion, for example through cytoreductive or bridging therapy \cite{park2018long, feuchtinger2024approaches, shahid2021bridging}.

Finally, we considered immune-privileged niches in which blasts can hide temporarily. We show that relapse risk increases significantly with the residence time within the niche. Short residence times enable elimination as blasts emerge before effector contraction, while longer times increase relapse probability as blasts reappear after effectors have diminished. Increasing effector persistence, for example by reducing the terminal effector loss rate $\delta$, raises the critical residence-time threshold below which elimination remains likely. Critically, the relevant form of persistence here is effector rather than memory persistence: niche-derived blasts emerge at numbers too low to drive substantial activation of the memory pool, so prevention of relapse depends on the sustained presence of effector CAR T-cells in circulation rather than the size of the memory reservoir.

Taken together, these results suggest that the clinical association between CAR T-cell persistence and durable remission in B-ALL reflects at least two distinct phenomena that the BEAM model can separate. Memory persistence sustains effector surveillance against slowly proliferating residual disease, and replenishes secondary effector waves against aggressive disease before the memory pool is exhausted. Effector persistence is what prevents relapse driven by blasts emerging stochastically from immune-privileged sites, since such blasts emerge at numbers too low to mount a substantial memory response in time. Our analysis does not commit to one mechanism being dominant: both forms of persistence are predicted to matter, and both fail through distinct mechanisms that the framework reproduces. A practical consequence is that clinical efforts to improve persistence may need to target different cellular subsets depending on the dominant relapse risk in a given patient.

Together, these results generate several experimentally testable predictions. First, among patients with comparable early MRD reduction, longer memory CAR T-cell persistence should preferentially reduce CD19-positive relapse driven by loss of immune surveillance. Second, reducing disease burden before infusion should reduce the risk of CD19-negative relapse by lowering the cumulative number of CD19-positive blast divisions before immune control is established. Third, relapse from protected or poorly accessible niches should depend more strongly on the duration of effector persistence than on memory-cell abundance alone, because isolated blasts exiting the niche must be killed before they expand sufficiently to trigger a broader memory response. Fourth, products or conditions that strongly favour memory formation may improve long-term control but could also allow a larger early tumour peak if effector generation is insufficient. These predictions could be tested using longitudinal measurements of MRD, CAR T-cell abundance, CAR T-cell phenotype, B-cell aplasia, and relapse antigen status in clinical cohorts, or using \textit{in vivo} models in which initial tumour burden, CAR T-cell product composition, and tissue accessibility can be varied experimentally.

Several limitations of the model and the available clinical data should be noted. First, the model assumes a well-mixed population; the validity of this assumption depends on whether blasts reside primarily in the bone marrow, blood, or perivascular niches, and spatial or agent-based extensions would be needed to relax it. Second, individual CAR T-cells do not experience functional exhaustion or saturation of cytotoxic capacity in our model; the finite division chain captures terminal loss of effector potential but not antigen-driven exhaustion~\cite{lai2025impact, dolina2021cd8}, and we expect this to cause the model to overestimate the ability of small effector populations to clear large blast burdens. Third, the tumour-burden-dependent fate switch through $B_{1/2}$ is a phenomenological idealisation rather than a directly measurable cellular threshold (\Cref{modeldev}); the precise functional form is not determined by data. Fourth, the CAR T-cell product is partitioned into three coarse functional compartments rather than the four phenotypes (na\"ive, central memory, effector memory, terminally differentiated) measured in FELIX~\cite{roddie2024obecabtagene}; a finer partition could be validated against future phenotype-resolved temporal data. Finally, calibration relies primarily on aggregate cohort kinetics from one trial; quantitative validation against per-patient trajectories from FELIX and other CD19 CAR T-cell trials remains an obvious next step. Despite these limitations, the BEAM model identifies specific hypotheses about the timing, phenotype and persistence of CAR T-cell responses that can be tested in clinical or \textit{in vivo} settings.

Beyond addressing these limitations, two clinical directions emerge as natural extensions. First, the model can be used to investigate the timing and intensity of pre-conditioning regimens, in particular bridging or cytoreductive therapy aimed at reducing $B_0$~\cite{feuchtinger2024approaches, shahid2021bridging} to reduce CD19-negative relapse risk. Second, the analysis of elimination after relapse suggested that a delayed second infusion of CAR T-cells, conditional on a sufficient remaining memory pool, could in principle drive elimination in a subset of patients, although this strategy presents significant toxicity concerns. Both directions would benefit from patient-specific re-calibration as longitudinal phenotype-resolved data become more widely available.

In summary, the BEAM model provides a quantitative framework in which the clinical link between CAR T-cell persistence and durable remission in B-ALL can be studied mechanistically. Returning to the question that motivated the work, the model suggests that memory and effector persistence are both required, but for distinct reasons: memory persistence sustains surveillance against low-burden or slow-growing residual disease, while effector persistence clears isolated blasts emerging stochastically from immune-privileged sites. The model also predicts a trade-off between immediate cytotoxicity and durable surveillance in product engineering, identifies initial tumour burden as a tractable lever for reducing antigen-negative relapse, and yields specific testable predictions about the relationship between phenotypic composition, kinetics, and relapse modality. Engineering CAR T-cells with an appropriate balance between early effector activity and long-term persistence, and tailoring this balance to individual disease characteristics, may reduce relapse risk and improve the durability of response.

\enlargethispage{20pt}

\ethics{This work did not require ethical approval from a human subject or animal welfare committee.}

\dataccess{All MATLAB code used to run simulations, generate figures and reproduce results is publicly available at \url{https://github.com/alexisfucl/mathematical_modelling_CART_immune_persistence}.}

\aucontribute{A.F.: conceptualization, formal analysis, investigation, methodology, software, validation, visualization, writing; B.J.W.: conceptualization, supervision, writing, M.A.P. conceptualization; K.M.P.: conceptualization, supervision, writing. All authors gave final approval for publication and agreed to be held accountable for the work performed therein.}

\competing{We declare we have no competing interests.}

\funding{A.F.’s work was supported by a RES scholarship from UCL. K.M.P. acknowledges funding from the NexTGen Cancer Grand Challenges partnership funded by Cancer Research UK (CGCATF-2021/100002) and the National Cancer Institute (CA278687-01) and The Mark Foundation for Cancer Research.}



\printbibliography
\appendix
\crefalias{section}{appendix}
\section{Hybrid simulation algorithm} \label{appA}

The model defined by \Cref{eq_blast_1}--\eqref{memory1} admits both a deterministic mean-field description, used when populations are large, and a stochastic reaction-network description, needed when population sizes become small enough for discrete fluctuations to matter. The reaction network is summarised in \Cref{tab:reaction_network}. The deterministic ODE model is recovered by replacing each reaction by its mean-field contribution, $\sum_j \nu_j \alpha_j(x)$, where $\nu_j$ is the stoichiometric update vector and $\alpha_j(x)$ is the reaction propensity. Logistic tumour growth is represented as two separate reactions: blast birth at rate $k_1 B$ and density-dependent blast death at rate $k_1 B^2/K$. The same convention is used when modelling antigen loss during blast division (\Cref{antigenescape}).

The algorithm proceeds in macro-steps of size $\Delta t$. At the start of each macro-step the state $x(t)$ is used to classify each reaction as deterministic or stochastic according to the criterion stated below. The deterministic reactions are integrated over $[t, t+\Delta t]$ under the provisional assumption that no stochastic reaction occurs, using MATLAB's \texttt{ode45} solver (an explicit Runge--Kutta (4,5) pair, the Dormand--Prince method). Along this provisional trajectory we evaluate the time-dependent propensities of the reactions currently classified as stochastic. For the inhomogeneous Poisson process with rate $\lambda(t)$, the time to the next event has distribution function
\begin{equation}
F(r) = 1 - \exp(-\Lambda(r)), \qquad \Lambda(r) = \int_t^{t+r} \lambda(s)\,ds, \qquad r > 0.
\end{equation}
Using the inverse-distribution method, we draw $u \sim U([0,1])$ and solve $F(r) = u$, giving $\Lambda(r) = -\log(1-u)$. Let $\mathcal{S}(t)$ denote the set of stochastic reactions at the start of the macro-step, and let $\alpha_j(s)$ be the propensity of reaction $R_j$ along the provisional deterministic trajectory. The time $t+\tau$ of the next discrete event satisfies
\begin{equation}\label{discretereac}
\int_t^{t+\tau} \sum_{j \in \mathcal{S}(t)} \alpha_j(s)\,ds = -\log(1-u),
\end{equation}
with $0 \le \tau \le \Delta t$. If the integrated stochastic propensity over the macro-step is smaller than $-\log(1-u)$, no stochastic event occurs in $[t, t+\Delta t]$ and the provisional ODE solution at $t+\Delta t$ is accepted. Otherwise a discrete reaction fires at $t+\tau$, and the conditional probability that reaction $i$ is the one to fire is
\begin{equation}
\mathbb{P}[j = i \mid T = t+\tau] = \frac{\alpha_i(t+\tau)}{\sum_{j \in \mathcal{S}(t)} \alpha_j(t+\tau)}, \qquad i \in \mathcal{S}(t).
\end{equation}
The stoichiometric jump of the chosen reaction is applied instantaneously to the deterministic state at $t+\tau$, and the algorithm restarts at $t+\tau$ with the reaction partition recalculated. It only remains to specify the partition criterion. Following~\cite{sto_sim}, reaction $R_j$ is treated deterministically at the start of a macro-step if both
\begin{enumerate}
\item the expected next reaction time $\tau_j = 1/\alpha_j(t)$ is smaller than a specified time increment $\delta t$, and
\item the number of individuals $X_i(t)$ for each species $i$ involved as a reactant in $R_j$ exceeds a specified threshold $\Lambda$.
\end{enumerate}
In our simulations we use $\delta t = 0.1$ day and $\Lambda = 10^{3}$. Reactions failing either condition are simulated stochastically during the macro-step. As the system evolves, the set of fast reactions changes dynamically: at low tumour burden, blast reactions become stochastic, and any CAR T-cell reaction whose propensity or reactant population falls below the threshold is also reclassified. The values were chosen to ensure good accuracy and efficiency. The qualitative results are insensitive to small changes around those values.

\begin{table}[H]
\centering
\caption{Reaction network used in the hybrid stochastic--deterministic simulations. Here, $E=\sum_{i=1}^{N}E_i$.}
\label{tab:reaction_network}
\begin{tabular}{@{}ccc@{}}
\toprule
\textbf{reaction} & \textbf{state update} & \textbf{propensity}\\
\midrule
$R_1$ & $B \rightarrow B+1$ & $k_1B$ \\
$R_2$ & $B \rightarrow B-1$ & $k_1B^2/K$ \\
$R_3$ & $B \rightarrow B-1$ & $k_2BE$ \\
$R_4$ & $M \rightarrow M-1,\; A \rightarrow A+1$ & $k_3MB$ \\
$R_5$ & $A \rightarrow A-1,\; M \rightarrow M+2$ & $k_4A\dfrac{B_{1/2}}{B_{1/2}+B}$ \\
$R_6$ & $A \rightarrow A-1,\; E_1 \rightarrow E_1+2$ & $k_4A\dfrac{B}{B_{1/2}+B}$ \\
$R_{6+i}$ & $E_i \rightarrow E_i-1$ & $\gamma E_i$ \\
 & $E_{i+1}\rightarrow E_{i+1}+2$ & $i=1,\ldots,N-1$ \\
$R_{N+6}$ & $E_N \rightarrow E_N-1$ & $\delta E_N$ \\
$R_{N+7}$ & $M \rightarrow M-1$ & $\epsilon M$ \\
\bottomrule
\end{tabular}
\end{table}

\section{Steady-state and stability analysis} \label{appB}

\subsection*{Dimensionless system}

With the scaling $\tau = k_1 t$, $b = B/K$, $e_i = (k_2/k_1) E_i$, $a = (k_2/k_1) A$, $m = (k_2/k_1) M$, and the dimensionless parameter groups $b_{1/2}$, $\eta$, $\rho$, $g$, $d$, $\tilde{\epsilon}$ defined in \Cref{bifurc}, \Cref{eq_blast_1}--\eqref{memory1} become
\begin{subequations} \label{dimless_system}
\begin{align}
\frac{\mathrm{d}b}{\mathrm{d}\tau} &= b(1-b) - b \sum_{i=1}^{N} e_i, &\qquad
\frac{\mathrm{d}e_N}{\mathrm{d}\tau} &= 2 g e_{N-1} - d e_N, \\
\frac{\mathrm{d}e_1}{\mathrm{d}\tau} &= 2\rho a \frac{b}{b_{1/2}+b} - g e_1, &\qquad
\frac{\mathrm{d}a}{\mathrm{d}\tau} &= \eta m b - \rho a, \\
\frac{\mathrm{d}e_i}{\mathrm{d}\tau} &= g(2 e_{i-1} - e_i),\quad i = 2, \ldots, N-1, &\qquad
\frac{\mathrm{d}m}{\mathrm{d}\tau} &= -\eta m b + 2 \rho a \frac{b_{1/2}}{b_{1/2}+b} - \tilde{\epsilon} m.
\end{align}
\end{subequations}

\subsection*{Steady states}

Setting all time derivatives in \eqref{dimless_system} to zero gives four steady states. The trivial states are $P_0 = (b^* = 0, \mathbf{0})$ and $P_1 = (b^* = 1, \mathbf{0})$. For non-trivial coexistence states, the activated-cell balance gives $a^* = \eta m^* b^*/\rho$. Substituting into the memory balance and assuming $m^* \neq 0$ yields a quadratic in $b^*$,
\begin{equation}\label{tocomp} 
\eta (b^*)^2 - (\eta b_{1/2} - \tilde{\epsilon}) b^* + \tilde{\epsilon} b_{1/2} = 0, \qquad \text{with solutions} \qquad b^*_{\pm} = \frac{\eta b_{1/2} - \tilde{\epsilon} \pm \sqrt{\left(\eta b_{1/2} - \tilde{\epsilon}\right)^2 - 4 \eta \tilde{\epsilon} b_{1/2}}}{2\eta}.
\end{equation}
The two positive coexistence states exist if and only if $\left(\eta b_{1/2} - \tilde{\epsilon}\right)^2 \geq 4 \eta \tilde{\epsilon} b_{1/2}$, which simplifies to
\begin{equation}
\tilde{\epsilon} \leq \frac{\eta b_{1/2}}{3 + 2\sqrt{2}}, \qquad \text{equivalently}, \qquad \epsilon \leq \frac{k_3 B_{1/2}}{3 + 2\sqrt{2}},
\end{equation}
recovering \eqref{foldbif}. The discriminant vanishes at $b^* = b_{1/2}(\sqrt{2}-1)$, and the two roots collide in a saddle-node (fold) bifurcation. In dimensional variables,
\begin{equation} \label{dimensional_coexistence}
B^*_{\pm} = \frac{k_3 B_{1/2} - \epsilon \pm \sqrt{\left(k_3 B_{1/2} - \epsilon\right)^2 - 4 k_3 B_{1/2} \epsilon}}{2 k_3}.
\end{equation}
The remaining coordinates of the coexistence states follow from the effector-chain balance. Setting $\mathrm{d}E_1/d\tau = 0$ gives $e_1^* = 2\rho a^* b^*/[g(b_{1/2}+b^*)]$; the recursion $\mathrm{d}E_i/d\tau = 0$ gives $e_i^* = 2^{i-1} e_1^*$ for $i = 2, \ldots, N-1$; and $\mathrm{d}E_N/d\tau = 0$ gives $e_N^* = (g/d)\,2^{N-1} e_1^*$. Combining with the blast balance $\sum_i e_i^* = 1 - b^*$ yields
\begin{equation} \label{Qdef}
e_1^* = \frac{1-b^*}{Q}, \qquad Q = 2^{N-1} - 1 + \frac{g}{d}\,2^{N-1},
\end{equation}
so the effector-chain total is fixed by the memory dynamics through $1 - b^*$ alone, and the per-compartment levels depend on $Q$, i.e. on $\gamma/\delta$ and $N$. For the baseline value $\gamma/\delta = 1.5$ and $N = 6$, $Q \approx 79$. The activated- and memory-cell levels at coexistence are then
\begin{subequations} \label{coexistence_coordinates}
\begin{align}
a^* &= \frac{g(b_{1/2}+b^*)}{2\rho b^*}\,\frac{1-b^*}{Q}, \qquad 
m^* = \frac{\rho a^*}{\eta b^*}.
\end{align}
\end{subequations}
A useful consequence is that the dimensional dormant blast burden $B^*_-$ in \eqref{dimensional_coexistence} depends only on $k_3$, $B_{1/2}$ and $\epsilon$; the effector-chain parameters $\gamma$, $\delta$ and $N$ affect the CAR T-cell levels required to sustain dormancy but not the blast level at which it is sustained.

\subsection*{Linear stability}

We now consider the linear stability of the steady states of Equation~(5). At the tumour-free state $P_0=(b^*=0,\mathbf{0})$, linearisation of the blast equation gives $\mathrm{d}b/\mathrm{d}\tau\sim b$, and hence the Jacobian has an eigenvalue $+1$. All remaining eigenvalues have negative real part: the effector-chain eigenvalues are $-g$ and $-d$, while the activated--memory subsystem has eigenvalues $-\rho$ and $-\tilde{\epsilon}$. Thus $P_0$ is a saddle. In particular, tumour elimination is not a stable deterministic equilibrium: an arbitrarily small positive blast population grows initially in the absence of an immune response.

At the tumour-escape state $P_1=(b^*=1,\mathbf{0})$, the blast direction contributes an eigenvalue $-1$. The effector chain contributes the eigenvalues $-g$ and $-d$. The remaining two eigenvalues are those of the activated--memory block
\begin{equation}
J_{AM}=
\begin{pmatrix}
-\rho & \eta\\
\dfrac{2\rho b_{1/2}}{1+b_{1/2}} & -(\eta+\tilde{\epsilon})
\end{pmatrix}.
\end{equation}
Its trace is
\begin{equation}
\operatorname{tr}(J_{AM})=-(\rho+\eta+\tilde{\epsilon})<0,
\end{equation}
while its determinant is
\begin{equation}
\det(J_{AM})=\rho\left[\tilde{\epsilon}+\eta\frac{1-b_{1/2}}{1+b_{1/2}}\right].
\end{equation}
Hence, for the biologically relevant regime $0<b_{1/2}<1$, we have $\det(J_{AM})>0$. The two eigenvalues of $J_{AM}$ therefore have negative real parts. Since all other eigenvalues are also negative, $P_1$ is linearly stable.

For the coexistence states $P_\pm$, the Jacobian is fully coupled and the characteristic polynomial does not yield a useful closed-form stability criterion for general $N$. At the fold, where $b^*=b_{1/2}(\sqrt{2}-1)$, one eigenvalue is zero, as expected for a saddle-node bifurcation. Away from the fold, we evaluate the Jacobian spectrum numerically over the parameter regimes considered in this study. We find that the upper coexistence branch $P_+$ has at least one eigenvalue with positive real part and is therefore unstable, whereas the lower branch $P_-$ has eigenvalues with negative real parts and is linearly stable over the parameter regimes considered. In parts of this region the dominant eigenvalues form a complex-conjugate pair, corresponding to damped oscillations towards the dormant coexistence state.

\section{Early-response analytical approximation of the cumulative blast burden} \label{appC}

The antigen-escape calculation in the main text depends on the cumulative CD19-positive blast burden, $\int_0^T B(t)\,\mathrm{d}t$. Here we derive an analytical approximation for this quantity during the initial treatment response. The reduction is based on two early-time assumptions: first, the blast burden is large compared with the phenotypic switching threshold, $B\gg B_{1/2}$, so that activated CAR T-cells divide predominantly into the effector lineage; second, the initial conversion of memory cells into activated cells occurs on a shorter timescale than the subsequent decay of the activated population. Once the resulting early-time effector response is prescribed, the blast equation is a non-autonomous logistic equation and can be solved exactly.

\subsection*{Activated-cell dynamics during the initial response}

The activated-cell equation is
\begin{equation}
\frac{\mathrm{d}A}{\mathrm{d}t}=k_3MB-k_4A\left(1-\frac{B}{B_{1/2}+B}\right)-k_4A\frac{B}{B_{1/2}+B}.
\end{equation}
Because the two $k_4$ terms sum exactly to $-k_4A$, this simplifies, without approximation, to
\begin{equation}\label{eq:A_exact_simplified}
\frac{\mathrm{d}A}{\mathrm{d}t}=k_3MB-k_4A.
\end{equation}
During the early high-burden phase, $B\gg B_{1/2}$, and hence
\begin{equation}
\frac{B}{B_{1/2}+B}\simeq 1, \qquad \frac{B_{1/2}}{B_{1/2}+B}\simeq 0.
\end{equation}
The memory equation therefore reduces to
\begin{equation}\label{eq:M_early}
\frac{\mathrm{d}M}{\mathrm{d}t}\simeq -(k_3B+\epsilon)M.
\end{equation}
On the short initial activation timescale we further take $B(t)\simeq B_0$, where $B_0=B(0)$. Define $\lambda_M:=k_3B_0+\epsilon$, then \Cref{eq:M_early} gives
\begin{equation}\label{eq:M_fast_solution}
M(t)\simeq M_0e^{-\lambda_M t},
\end{equation}
where $M_0=M(0)$. Substituting \Cref{eq:M_fast_solution} into \Cref{eq:A_exact_simplified} gives
\begin{equation}\label{eq:A_forced}
\frac{\mathrm{d}A}{\mathrm{d}t}+k_4A=k_3B_0M_0e^{-\lambda_Mt}.
\end{equation}
Multiplication by the integrating factor $e^{k_4t}$ yields
\begin{equation}
\frac{\mathrm{d}}{\mathrm{d}t}\left(e^{k_4t}A(t)\right)=k_3B_0M_0e^{(k_4-\lambda_M)t}.
\end{equation}
For $\lambda_M\neq k_4$, integration from $0$ to $t$ gives
\begin{equation}\label{eq:A_fast_solution}
A(t)\simeq A_0e^{-k_4t}+\frac{k_3B_0M_0}{\lambda_M-k_4}\left(e^{-k_4t}-e^{-\lambda_Mt}\right),
\end{equation}
where $A_0=A(0)$. In the exceptional case $\lambda_M=k_4$, the corresponding solution is $A(t)\simeq e^{-k_4t}(A_0+k_3B_0M_0t)$. For the parameter regime of interest the initial activation rate is fast relative to activated-cell loss, so that $\lambda_M\gg k_4$. After the short activation transient, $t\gg \lambda_M^{-1}$, the term proportional to $e^{-\lambda_Mt}$ in \Cref{eq:A_fast_solution} is negligible and
\begin{equation}\label{eq:A_outer}
A(t)\simeq A_*e^{-k_4t}, \qquad A_*:=A_0+\frac{k_3B_0M_0}{\lambda_M-k_4}.
\end{equation}
If, in addition, $k_3B_0\gg k_4,\epsilon$, then $A_*\simeq A_0+M_0$. Thus $A_*$ is the effective activated-cell amplitude after the rapid initial activation transient; it is not, in general, equal to the literal initial value $A(0)$.

\subsection*{Linear effector-chain response}

Under the same high-burden approximation, $B/(B_{1/2}+B)\simeq 1$, the effector equations become
\begin{subequations}\label{eq:E_chain_early}
\begin{align}
\frac{\mathrm{d}E_1}{\mathrm{d}t} &= 2k_4A-\gamma E_1, \\
\frac{\mathrm{d}E_i}{\mathrm{d}t} &= 2\gamma E_{i-1}-\gamma E_i, \qquad i=2,\ldots,N-1, \\
\frac{\mathrm{d}E_N}{\mathrm{d}t} &= 2\gamma E_{N-1}-\delta E_N.
\end{align}
\end{subequations}
Define the effector-state vector
\begin{equation}
\mathbf{E}(t):=\begin{pmatrix}E_1(t)&E_2(t)&\cdots&E_N(t)\end{pmatrix}^{\!T},
\end{equation}
the first standard basis vector
\begin{equation}
\mathbf{e}_1:=\begin{pmatrix}1&0&\cdots&0\end{pmatrix}^{\!T}\in\mathbb{R}^N,
\end{equation}
and the $N\times N$ effector-chain matrix
\begin{equation}\label{eq:QE_matrix}
\mathbf{Q}_E:=
\begin{pmatrix}
-\gamma & 0 & 0 & \cdots & 0 \\
2\gamma & -\gamma & 0 & \cdots & 0 \\
0 & 2\gamma & -\gamma & \ddots & \vdots \\
\vdots & \ddots & \ddots & \ddots & 0 \\
0 & \cdots & 0 & 2\gamma & -\delta
\end{pmatrix}.
\end{equation}
Thus $(\mathbf{Q}_E)_{ii}=-\gamma$ for $i=1,\ldots,N-1$, $(\mathbf{Q}_E)_{NN}=-\delta$, $(\mathbf{Q}_E)_{i,i-1}=2\gamma$ for $i=2,\ldots,N$, and all other entries vanish. The notation $\mathbf{Q}_E$ is used here to distinguish this matrix from the scalar quantity $Q$ defined in \Cref{Qdef}. With these definitions, \Cref{eq:E_chain_early} is
\begin{equation}\label{eq:E_matrix_ode}
\frac{\mathrm{d}\mathbf{E}}{\mathrm{d}t}=\mathbf{Q}_E\mathbf{E}+2k_4A(t)\mathbf{e}_1.
\end{equation}
Let $\mathbf{E}^{(0)}:=\mathbf{E}(0)$ denote the vector of initially infused effector cells. The variation-of-constants formula gives
\begin{equation}\label{eq:E_variation_constants}
\mathbf{E}(t)=e^{\mathbf{Q}_Et}\mathbf{E}^{(0)}+2k_4\int_0^t e^{\mathbf{Q}_E(t-s)}\mathbf{e}_1A(s)\,\mathrm{d}s,
\end{equation}
where $e^{\mathbf{Q}_Et}$ is the matrix exponential. The total effector population is
\begin{equation}
E(t):=\sum_{i=1}^{N}E_i(t)=\mathbf{1}^{T}\mathbf{E}(t),
\end{equation}
where $\mathbf{1}:=(1,\ldots,1)^T\in\mathbb{R}^N$. Hence
\begin{equation}\label{eq:E_total_convolution}
E(t)=\underbrace{\mathbf{1}^{T}e^{\mathbf{Q}_Et}\mathbf{E}^{(0)}}_{E_{\mathrm{init}}(t)}+\underbrace{2k_4\int_0^t \mathbf{1}^{T}e^{\mathbf{Q}_E(t-s)}\mathbf{e}_1A(s)\,\mathrm{d}s}_{E_A(t)}.
\end{equation}
The first term is the contribution from effector cells already present at infusion, while the second is the contribution generated from the activated-cell compartment. Substituting the exponential approximation \Cref{eq:A_outer} into \Cref{eq:E_variation_constants} gives
\begin{equation}
\mathbf{E}(t)=e^{\mathbf{Q}_Et}\mathbf{E}^{(0)}+2k_4A_*\int_0^t e^{\mathbf{Q}_E(t-s)}\mathbf{e}_1e^{-k_4s}\,\mathrm{d}s.
\end{equation}
Since the scalar matrix $k_4\mathbf{I}$ commutes with $\mathbf{Q}_E$, where $\mathbf{I}$ is the $N\times N$ identity matrix,
\begin{equation}
e^{\mathbf{Q}_E(t-s)}e^{-k_4s}=e^{\mathbf{Q}_Et}e^{-(\mathbf{Q}_E+k_4\mathbf{I})s}.
\end{equation}
Provided $\mathbf{Q}_E+k_4\mathbf{I}$ is invertible, equivalently $k_4\neq\gamma$ and $k_4\neq\delta$, we therefore obtain
\begin{equation}\label{eq:E_matrix_closed}
\mathbf{E}(t)=e^{\mathbf{Q}_Et}\mathbf{E}^{(0)}+2k_4A_*(\mathbf{Q}_E+k_4\mathbf{I})^{-1}\left(e^{\mathbf{Q}_Et}-e^{-k_4t}\mathbf{I}\right)\mathbf{e}_1.
\end{equation}
Consequently,
\begin{equation}\label{eq:E_total_closed}
E(t)=\mathbf{1}^{T}e^{\mathbf{Q}_Et}\mathbf{E}^{(0)}+2k_4A_*\mathbf{1}^{T}(\mathbf{Q}_E+k_4\mathbf{I})^{-1}\left(e^{\mathbf{Q}_Et}-e^{-k_4t}\mathbf{I}\right)\mathbf{e}_1.
\end{equation}
If $k_4=\gamma$ or $k_4=\delta$, the integral representation \Cref{eq:E_variation_constants} remains valid and the corresponding limiting expression contains polynomial--exponential terms. Equation~\eqref{eq:E_total_convolution} also makes precise the interpretation of the effector response as a delayed and amplified version of the activated-cell dynamics. Define the impulse-response kernel
\begin{equation}\label{eq:effector_kernel}
h(u):=2k_4\mathbf{1}^{T}e^{\mathbf{Q}_Eu}\mathbf{e}_1, \qquad u\geq 0.
\end{equation}
Then
\begin{equation}\label{eq:E_distributed_delay}
E_A(t)=\int_0^t h(u)A(t-u)\,\mathrm{d}u.
\end{equation}
Thus the finite division chain generates a distributed delay rather than an exact single fixed delay. If $A$ varies sufficiently slowly compared with the width of $h$, a first-moment approximation gives
\begin{equation}\label{eq:E_fixed_delay_approx}
E_A(t)\simeq \widetilde{e}\,A(t-\tau),
\end{equation}
where
\begin{equation}\label{eq:effector_gain_delay}
\widetilde{e}:=\int_0^\infty h(u)\,\mathrm{d}u=-2k_4\mathbf{1}^{T}\mathbf{Q}_E^{-1}\mathbf{e}_1, \qquad \tau:=\frac{\int_0^\infty uh(u)\,\mathrm{d}u}{\int_0^\infty h(u)\,\mathrm{d}u}=-\frac{\mathbf{1}^{T}\mathbf{Q}_E^{-2}\mathbf{e}_1}{\mathbf{1}^{T}\mathbf{Q}_E^{-1}\mathbf{e}_1}.
\end{equation}
The amplification factor may also be written explicitly as
\begin{equation}\label{eq:effector_gain_explicit}
\widetilde{e}=k_4\left(\frac{2^N-2}{\gamma}+\frac{2^N}{\delta}\right).
\end{equation}
For quantitative calculations below we retain the matrix solution \Cref{eq:E_total_closed}; \Cref{eq:E_fixed_delay_approx} is used only to interpret the division chain as an amplified, delayed response to $A$.

\subsection*{Solution of the blast equation for a prescribed effector response}

Once $E(t)$ is prescribed, the blast equation may be written
\begin{equation}\label{eq:B_nonauto_logistic}
\frac{\mathrm{d}B}{\mathrm{d}t}=\left[k_1-k_2E(t)\right]B-\frac{k_1}{K}B^2.
\end{equation}
Define
\begin{equation}\label{eq:g_psi_def}
g(t):=k_1-k_2E(t), \qquad \Psi(t):=\int_0^t g(s)\,\mathrm{d}s=k_1t-k_2\int_0^tE(s)\,\mathrm{d}s.
\end{equation}
Introducing $Y(t):=1/B(t)$ transforms \Cref{eq:B_nonauto_logistic} into the linear equation
\begin{equation}\label{eq:Y_linear}
\frac{\mathrm{d}Y}{\mathrm{d}t}+g(t)Y=\frac{k_1}{K}.
\end{equation}
Indeed,
\begin{equation}
\frac{\mathrm{d}Y}{\mathrm{d}t}=-\frac{1}{B^2}\frac{\mathrm{d}B}{\mathrm{d}t}=-\frac{g(t)}{B}+\frac{k_1}{K}=-g(t)Y+\frac{k_1}{K}.
\end{equation}
The integrating factor for \Cref{eq:Y_linear} is $e^{\Psi(t)}$, since $\Psi'(t)=g(t)$. Hence
\begin{equation}
\frac{\mathrm{d}}{\mathrm{d}t}\left(e^{\Psi(t)}Y(t)\right)=\frac{k_1}{K}e^{\Psi(t)}.
\end{equation}
Integrating from $0$ to $t$, using $Y(0)=1/B_0$ and $\Psi(0)=0$, gives
\begin{equation}
Y(t)=e^{-\Psi(t)}\left[\frac{1}{B_0}+\frac{k_1}{K}\int_0^t e^{\Psi(s)}\,\mathrm{d}s\right].
\end{equation}
Taking the reciprocal therefore yields
\begin{equation}\label{eq:B_early_solution}
B(t)=\frac{B_0e^{\Psi(t)}}{1+\dfrac{k_1B_0}{K}\int_0^t e^{\Psi(s)}\,\mathrm{d}s}, \qquad \Psi(t)=k_1t-k_2\int_0^tE(s)\,\mathrm{d}s.
\end{equation}
This expression is exact for the blast equation once $E(t)$ has been prescribed; the approximation enters through the early-response expression used for $E(t)$. For completeness, the cumulative effector exposure appearing in $\Psi$ can be obtained directly from \Cref{eq:E_total_closed}. Since $\mathbf{Q}_E$ is invertible,
\begin{equation}
\int_0^t e^{\mathbf{Q}_Es}\,\mathrm{d}s=\mathbf{Q}_E^{-1}\left(e^{\mathbf{Q}_Et}-\mathbf{I}\right).
\end{equation}
It follows that
\begin{align}\label{eq:effector_exposure_closed}
\int_0^tE(s)\,\mathrm{d}s={}&\mathbf{1}^{T}\mathbf{Q}_E^{-1}\left(e^{\mathbf{Q}_Et}-\mathbf{I}\right)\mathbf{E}^{(0)} \notag\\
&+2k_4A_*\mathbf{1}^{T}(\mathbf{Q}_E+k_4\mathbf{I})^{-1}\left[\mathbf{Q}_E^{-1}\left(e^{\mathbf{Q}_Et}-\mathbf{I}\right)-\frac{1-e^{-k_4t}}{k_4}\mathbf{I}\right]\mathbf{e}_1.
\end{align}
Thus $\Psi(t)$ is explicit up to evaluation of matrix exponentials.

\subsection*{Cumulative blast burden}

The quantity relevant to the antigen-escape calculation is
\begin{equation}
J(T):=\int_0^T B(t)\,\mathrm{d}t.
\end{equation}
Define
\begin{equation}\label{eq:D_def}
D(t):=1+\frac{k_1B_0}{K}\int_0^t e^{\Psi(s)}\,\mathrm{d}s.
\end{equation}
Then \Cref{eq:B_early_solution} is $B(t)=B_0e^{\Psi(t)}/D(t)$, while differentiation of \Cref{eq:D_def} gives
\begin{equation}
D'(t)=\frac{k_1B_0}{K}e^{\Psi(t)}.
\end{equation}
Therefore
\begin{equation}
\frac{D'(t)}{D(t)}=\frac{k_1}{K}B(t),
\end{equation}
and hence
\begin{equation}
B(t)=\frac{K}{k_1}\frac{\mathrm{d}}{\mathrm{d}t}\log D(t).
\end{equation}
Integrating from $0$ to $T$ and using $D(0)=1$ gives
\begin{equation}\label{eq:cumulative_blast_early}
\int_0^T B(t)\,\mathrm{d}t=\frac{K}{k_1}\log\left[1+\frac{k_1B_0}{K}\int_0^T e^{\Psi(s)}\,\mathrm{d}s\right].
\end{equation}
Equations~\eqref{eq:B_early_solution} and \eqref{eq:cumulative_blast_early} are the expressions quoted in the main text. They show explicitly that the initial tumour trajectory and cumulative blast burden are controlled by the cumulative effector exposure $\int_0^tE(s)\,\mathrm{d}s$. The analytical reduction therefore provides a mechanistic interpretation of the parameter dependencies in Figure~6: parameters that alter the strength or timing of the initial effector wave change $\Psi(t)$ and thereby the cumulative number of CD19-positive blast-cell divisions available for antigen loss.

\subsection*{Domain of validity}

The reduction above is intended as an early-response approximation rather than a uniformly valid solution of the full BEAM system. In particular, the approximations $B\gg B_{1/2}$ and $B\simeq B_0$ are used only to resolve the rapid initial memory-to-activated-cell transient, after which the exponentially decaying activated-cell approximation is propagated through the linear effector chain. The resulting formulae are expected to be most accurate while the initial effector wave dominates and before low-burden memory feedback or secondary immune responses become important. For parameter values for which these assumptions are not satisfied, the full system should be used.

\end{document}